\documentclass[letterpaper]{mn2e}

\usepackage{graphicx}
\def\simgt{\lower 2pt \hbox{$\, \buildrel {\scriptstyle >}\over{\scriptstyle \sim}\,$}}
\def\simlt{\lower 2pt \hbox{$\, \buildrel {\scriptstyle <}\over{\scriptstyle \sim}\,$}}
\newcommand\ion[2]{#1$\,${\tt #2}}
\newcommand{\apj}{ApJ}
\newcommand{\apjs}{ApJS}
\newcommand{\apjl}{ApJL}
\newcommand{\aap}{A{\&}A}

\newcommand{\mnras}{MNRAS}
\newcommand{\aj}{AJ}
\newcommand{\araa}{ARAA}

\newcommand{\procspie}{Proc.~SPIE}
\newcommand{\nat}{Nature}
\newcommand{\nodata}{...}

\title[BAL variability in RLQs]{Broad Absorption Line Variability in Radio-Loud Quasars}

\author[Welling, Miller, Brandt, Capellupo, \& Gibson]{C.~A.~Welling,$^{1}$ B.~P.~Miller,$^{2,3}$
W.~N.~Brandt,$^{4,5}$ D.~M.~Capellupo,$^{6,7}$ \&~R.~R.~Gibson$^{8}$\\
$^{1}$Department of Physics and Astronomy, Dickinson
  College, Carlisle, PA 17013\\
$^{2}$Department of Astronomy, University of Michigan, 500
  Church Street, Ann Arbor, MI 48109\\
$^{3}$Department of Physics and Astronomy, Macalester College, 
  1600 Grand Ave, Saint Paul, MN 55105\\
$^{4}$Department of Astronomy and Astrophysics, The
  Pennsylvania State University, 525 Davey Laboratory, University
  Park, PA 16802\\
$^{5}$Institute for Gravitation and the Cosmos, The
  Pennsylvania State University, University Park, PA 16802\\
$^{6}$Department of Astronomy, University of Florida, 211
  Bryant Space Science Center, Gainesville, FL 32611\\
$^{7}$School of Physics and Astronomy, Tel Aviv University,
  Tel Aviv 69978, Israel\\
$^{8}$Department of Astronomy, University of Washington,
  Box 351580, Seattle, WA 98195}

\begin{document}

\maketitle

\begin{abstract}

We investigate \ion{C}{IV} broad absorption line (BAL) variability
within a sample of 46 radio-loud quasars (RLQs), selected from
SDSS/FIRST data to include both core-dominated (39) and lobe-dominated
(7) objects. The sample consists primarily of high-ionization BAL
quasars, and a substantial fraction have large BAL velocities or
equivalent widths; their radio luminosities and radio-loudness values
span $\sim$2.5 orders of magnitude. We have obtained 34 new
Hobby-Eberly Telescope (HET) spectra of 28 BAL RLQs to compare to
earlier SDSS data, and we also incorporate archival coverage
(primarily dual-epoch SDSS) for a total set of 78 pairs of equivalent
width measurements for 46 BAL RLQs, probing rest-frame timescales of
$\sim$80--6000~d (median 500~d). In general, only modest changes in
the depths of segments of absorption troughs are observed, akin to
those seen in prior studies of BAL RQQs. Also similar to previous
findings for RQQs, the RLQs studied here are more likely to display
BAL variability on longer rest-frame timescales. However, typical
values of $|{\Delta}EW|$ and $|{\Delta}EW|/{\langle}EW{\rangle}$ are
$\sim$40$\pm20$\% lower for BAL RLQs when compared with those of a
timescale-matched sample of BAL RQQs. Optical continuum variability is
of similar amplitude in BAL RLQs and BAL RQQs; for both RLQs and RQQs,
continuum variability tends to be stronger on longer timescales. BAL
variability in RLQs does not obviously depend upon their radio
luminosities or radio-loudness values, but we do find tentative
evidence for greater fractional BAL variability within lobe-dominated
RLQs. Enhanced BAL variability within more edge-on (lobe-dominated)
RLQs supports some geometrical dependence to the outflow structure.

\end{abstract}

\begin{keywords}
galaxies: active --- galaxies: jets --- quasars: absorption lines 
\end{keywords}

\section{Introduction}

Accretion in quasars appears to lead naturally to the formation of
outflows that may regulate supermassive black hole growth and provide
feedback to the host galaxy (e.g., Arav et al.~2013 and references
therein), potentially helping to quench star formation. In radio-quiet
quasars (RQQs) such outflows are most readily apparent as broad
absorption lines (BALs; Weymann et al.~1981, 1991) found
blueward\footnote{A rare handful of quasars display redshifted BALs,
  perhaps from an infall or a rotating outflow (Hall et al.~2013).} of
UV emission lines; these features can occur at a wide range of
velocities (to greater than 0.1$c$) and are observed in 10--20\% of
optically selected quasars (e.g., Hewett \& Foltz 2003). The
``orientation'' model hypothesizes that BALs are common to RQQs but
only apparent over a limited range of inclinations to the line of
sight. (While successful at explaining many observed properties of BAL
and non-BAL quasars, this simple model may not capture the full
physical complexity of outflow generation and structure.) The
prevalence of velocity structure within \ion{C}{IV} BALs that matches
the Ly$\alpha$--N~V velocity offset (Weymann et al.~1991; Arav \&
Begelman 1994) indicates that these BAL outflows are radiatively
accelerated, as does the correlation between maximum outflow velocity
and UV luminosity (Laor \& Brandt 2002; Ganguly et
al.~2007). Simulations (e.g., Proga et al.~2000) demonstrate that
winds can be driven off a classical accretion disk, with interior
``shielding gas'' (Murray et al.~1995) preventing overionization and
likely accounting for \hbox{X-ray} absorption in BAL QSOs (e.g.,
Gallagher et al.~2006; see also Gibson et al.~2009a and Wu et al.~2010
for discussion of mini-BALs). Observational evidence favoring the
disk-wind model includes the relatively high degree of polarization
among BAL quasars in general and in BAL troughs in particular (e.g.,
Ogle et al.~1999; Young et al.~2007; DiPompeo et al.~2011) and the
similarity of the spectral energy distributions of BAL and non-BAL
quasars (e.g., Willott et al.~2003; Gallagher et al.~2007; but see
also DiPompeo et al.~2013). On the other hand, BAL quasars have been
argued to accrete at particularly high Eddington ratios (e.g., Ganguly
et al.~2007), as inferred based on apparent [O~III] weakness (Yuan \&
Wills 2003). Quasars also possessing low-ionization BALs (LoBALs, in
contrast to the more common high-ionization only HiBALs) in particular
tend toward weak [O~III] and seemingly lack [Ne~V] (e.g., Zhang et
al.~2010). After accounting for intrinsic absorption, Luo et
al.~(2013) estimate that 17--40\% of BAL quasars are still
\hbox{X-ray} weak, and suggest that \hbox{X-ray} weak quasars may more
easily launch outflows (due to reduced overionization) with
potentially large covering factors.

An initial lack of detected BALs among radio-loud quasars (RLQs) was
interpreted to indicate that jets and BALs were mutually exclusive
(e.g., Stocke et al.~1992). This paradigm was challenged by a series
of discoveries of individual BAL RLQs (e.g., Becker et al.~1997;
Brotherton et al.~1998; Wills et al.~1999; Gregg et al.~2000) and then
undermined by the identification of a population of BAL RLQs (e.g.,
Becker et al.~2000, 2001; Menou et al.~2001; Shankar et al.~2008),
mostly detected in the VLA 1.4~GHz FIRST survey (Becker et al.~1995)
with systematic optical spectroscopic coverage obtained by the FIRST
Bright Quasar Survey (FBQS; White et al.~2000) and the Sloan Digital
Sky Survey (SDSS; York et al.~2000). Several BAL RLQs display radio
spectral and/or morphological properties similar to those of compact
steep spectrum (CSS) or GHz-peaked spectrum (GPS) radio sources, which
are commonly presumed to be young (e.g., Stawarz et al.~2008),
although in general BAL RLQ radio morphologies do not require youth
(Bruni et al.~2013). Additionally, dust-reddened quasars (plausibly
newly active; Urrutia et al.~2008; Glikman et al.~2012) appear more
likely to host low-ionization BALs (Urrutia et al.~2009), suggesting
that at least ``LoBALs'' may be linked to source age rather than
inclination. These observations, in concert with the remaining
scarcity of BALs within strongly radio-loud and lobe-dominated
objects, have revived alternative ``evolutionary'' models (Gregg et
al.~2006) that associate BALs with emerging young quasars clearing
their kpc-scale environment through outflows spanning equatorial
through polar (e.g., Zhou et al.~2006) latitudes (though a purely
evolutionary model requires fine-tuning to match observations; Shankar
et al.~2008). The reality may lie between a stark
orientation/evolution dichotomy, with some types of quasars more able
to host winds that themselves have a range of covering factors
(Richards et al.~2011; DiPompeo et al.~2012, 2013). In any event, it
is currently unclear whether BALs in RLQs have a similar physical
origin to those in RQQs, or indeed whether BALs in RLQs are even a
homogeneous class.

Variability studies provide one method of assessing BAL structure, and
they can potentially constrain the location and dynamics of the UV
absorber. In principle, BAL variability could be induced through an
alteration in the ionization parameter as a result of fluctuations in
the incident flux (e.g., Trevese et al.~2013); the variability
timescale then constrains the absorber density (Netzer et al.~2002)
and/or distance (Narayanan et al.~2004). However, this is unlikely to
be the dominant mechanism for cases in which the \ion{C}{IV}
variability is confined to a restricted velocity segment within the
full BAL absorption trough, which indeed constitute the majority of
observed variability behavior (e.g., Gibson et al.~2008; Capellupo et
al.~2013; see also discussion of saturated \ion{C}{IV} outflows in the
latter). Alternatively, depth changes within BAL profiles may
plausibly arise from dynamical restructuring of the absorber along the
line of sight, with the variability timescale providing estimates of
crossing speeds and location (Risaliti et al.~2002; Capellupo et
al.~2013). Simulations suggest that although the shielding component
of the wind can be dynamic (Sim et al.~2012; see also observational
X-ray results from Saez et al.~2012), synthesized absorption profiles
are relatively constant at lower velocities whereas variabilty becomes
more pronounced at higher velocities, which correspond to
well-shielded material streaming to larger distances (Proga et
al.~2012). Such disk azimuthal asymmetries can potentially link
variability, at differing amplitudes, across multiple velocity
components (Filiz Ak et al.~2012). In general, dynamical wind outflow
models can produce extremely complex behavior (e.g., Giustini \& Proga
2012).

Broad absorption line variability within radio-quiet quasars has now
been characterized through several studies, many of which make use of
SDSS data for one or more epochs of coverage. BALs in RQQs often show
minor depth changes within narrow portions of troughs (Barlow 1993;
Lundgren et al.~2007; Gibson et al.~2008, 2010; Capellupo et al.~2011;
hereafter B93, L07, G10, and C11, respectively). Variability is
perhaps more common within shallower or higher-velocity BALs (L07;
C11) which are occasionally even observed to disappear completely
(Filiz Ak et al.~2012). BALs of greater equivalent width ($EW$) tend
to show greater absolute changes in $EW$, and BALs spanning a wider
velocity range tend to show variability within a larger absolute
subset of velocity bins (Gibson et al.~2008, their Figure 9 and
Equation 1, respectively). However, the absolute value of the
fractional change in $EW$ is greater in BALs of lower $EW$ (L07), and
similarly within a given velocity segment shallower absorption
increases the likelihood of variability (C11). Acceleration of BALs is
rarely observed (e.g., Hall et al.~2007; Gibson et al.~2008). Changes
in velocity width can sometimes transition features out of (or into)
formal BAL classification (e.g., Rodr{\'{\i}}guez Hidalgo et al.~2013;
Misawa et al.~2005; see also discussion in Gibson et al.~2009a),
indicative of a connection between narrow-absorption line or mini-BAL
troughs and the more extreme BALs. While BAL variability is not
necessarily monotonic, in general BALs in RQQs tend to vary more often
and more strongly on longer timescales (Gibson et al.~2008; G10; C11),
although variability on only rest-frame $\sim$8--10~d has been seen
(Capellupo et al.~2013).

These results provide a baseline for RQQ BAL variability, but to date
there has not been a systematic survey of BAL variability within
RLQs\footnote{A preliminary sketch of some portions of this paper is
  given in Miller et al.~(2012). Filiz Ak et al.~(2013) study BAL
  quasars in SDSS and briefly compare RQQs to a small set of RLQs.}
and so no statistical comparison has been possible. This work conducts
such a study through measurement of \ion{C}{IV} absorption at multiple
epochs in a large sample of RLQs. We have obtained 34 new spectra for
28 BAL RLQs, primarily selected as such from FIRST/SDSS data, with the
Hobby-Eberly Telescope (HET; Ramsey et al.~1998) Low-Resolution
Spectrograph (LRS; Hill et al.~1998). This sample was chosen to cover
a wide range in radio-loudness and luminosity and also BAL velocity
and equivalent width. BAL variability is assessed through a comparison
of the HET/LRS spectra to the earlier SDSS spectra. We also
incorporate BAL variability measurements obtained for 18 additional
RLQs with two (or more) SDSS or archival spectra available. Together,
the 46 RLQs have 78 pairs of BAL equivalent width measurements,
probing rest-frame timescales of $\sim$80--6000~d (median 800~d).

\begin{figure}
\includegraphics[scale=0.42]{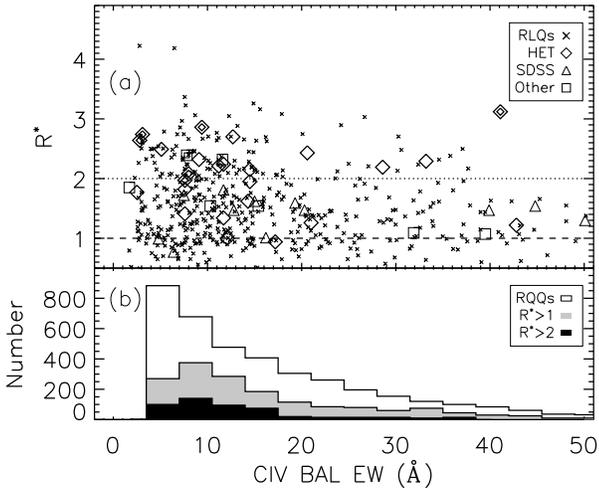} 
\caption{\small (a) Radio-loudness plotted versus \ion{C}{IV} BAL
  EW. The BAL RLQs studied here from HET/SDSS, SDSS/SDSS, and other
  archival pairs of spectra are plotted as diamonds, triangles, and
  squares, respectively. Lobe-dominated RLQs are marked with double
  symbols.  Additional BAL RLQs, identified from the G09 BAL catalog
  matched to FIRST data, are also shown (crosses). The dashed/dotted
  lines show increasingly restrictive cuts in $R^{*}$. (b)
  Distribution of \ion{C}{IV} BAL EW for RQQs and groups of RLQs.}
\end{figure}

This paper is organized as follows: $\S$2 describes the selection and
radio and BAL characteristics of the sample, $\S$3 quantifies BAL
variability, $\S$4 compares to results for BAL RQQs and investigates
dependencies upon continuum variability and radio properties, and
$\S$5 summarizes and concludes. We use positive values of equivalent
width (given in units of rest-frame~\AA) to quantify BAL absorption
strength, and express changes in equivalent width such that a positive
difference corresponds to the BAL deepening between observations. A
standard cosmology with $H_{\rm 0}=70$~km~s$^{-1}$~Mpc$^{-1}$,
${\Omega}_{\rm M}=0.3$, and ${\Omega}_{\rm \Lambda}=0.7$ is assumed
throughout. Monochromatic luminosities are given in units of
erg~s$^{-1}$~Hz$^{-1}$ and expressed as logarithms, with ${\ell}_{\rm
r}$ and ${\ell}_{\rm uv}$ determined at rest-frame 5~GHz and 2500~\AA,
respectively. Unless otherwise noted, errors are quoted at
1$\sigma$. Object names are given as SDSS J2000 and taken from the DR7
Quasar Catalog of Schneider et al.~(2010; see also Schneider et
al.~2005, 2007). 

\begin{figure}
\includegraphics[scale=0.42]{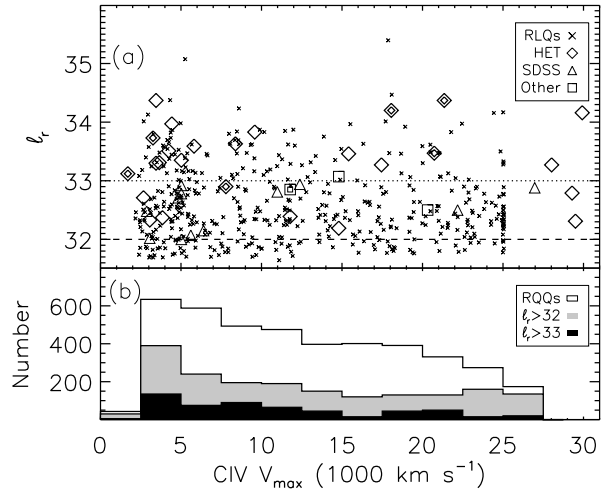}
\caption{\small (a) Radio luminosity plotted versus \ion{C}{IV}
  maximum outflow velocity, with symbols as in Figure~1. Note that the
  maximum outflow velocity is restricted to 25000~km~s$^{-1}$ in
  G09. The dashed/dotted lines show increasingly restrictive cuts in
  ${\ell}_{\rm r}$. (b) Distribution of $v_{\rm max}$ for RQQs and for
  RLQs with cuts as above. }
\end{figure}

\section{Sample Characteristics}

General optical, radio, and (longest separation) absorption properties
of the 46 BAL RLQs studied here are listed in Table~1. We restrict
consideration of absorption variability to the \ion{C}{IV} line. The
sample consists primarily of HiBALs; only 5/46 RLQs are known LoBALs
(although for RLQs with $z>2.1$ the \ion{Mg}{II} line is redshifted
out of SDSS coverage). The objects consequently have redshifts
$1.7<z<4$ so as to be accessible to ground-based telescopes, except
for 100726.10+124856.2 (PG~1004+130; $z=0.241$) which has {\it IUE\/}
and {\it HST\/} coverage.

\subsection{Sample selection}

The 28 objects observed with the HET were identified as BAL quasars in
the SDSS DR3 catalog of Trump et al.~(2006) and the SDSS DR5 catalog
of Gibson et al.~(2009b; hereafter G09). (The one exception to the HET
targeting of SDSS quasars is 101614.26+520915.7, discovered and
described in Gregg et al.~2000, which we include to increase the size
of our subsample of lobe-dominated BAL RLQs.) One of the 28 HET
targets is listed in Trump et al.~(2006) but not also identified as a
BAL quasar in Gibson et al.~(2009b): J074610.50$+$230710.7 has two
deep adjacent absorption troughs (in both \ion{C}{IV} and \ion{Si}{IV}),
which if considered together reach BAL width
($>2000$~km~s$^{-1}$). RLQs (defined as in $\S$2.2) were identified
through cross-matching to the FIRST survey catalog, with all radio
components within 90$''$ initially retained and then visually
classified as either lobes (e.g., an approximately symmetric pair
about the optical source) or else as background (e.g., with an SDSS
counterpart, or an unrelated intruding double). HET targets were
selected to have $1.7<z<4.0$ to permit optical access to the
\ion{C}{IV} region. The HET targets span a wide range in radio and BAL
properties, and as demonstrated in $\S$2.2 our full sample is
representative of the parent population of BAL RLQs with respect to
BAL equivalent width or maximum velocity, and with respect to radio
luminosity or radio loudness. For feasibility purposes brighter
targets were preferred, and as the redshift coverage is representative
this results in the SDSS/HET objects having somewhat greater optical
luminosities than the parent population of BAL RLQs.

An additional 18 BAL RLQs with multiple SDSS or archival spectra are
included in our full sample. Most of these objects were identified as
BAL RLQs through cross-matching the G09 catalog with FIRST as for the
HET sample. The separation between spectra was required to exceed
rest-frame 15~d. The SDSS/SDSS objects are observed on relatively
short rest-frame timescales, similar to the 20--120~d range probed by
the multi-epoch SDSS study of BAL RQQs conducted by Lundgren et
al.~(2007). A few of the SDSS/HET objects also have multiple SDSS
spectra, and we also quantify their shorter timescale variability. A
handful of these additional BAL RLQs were selected from the literature
(e.g., L07; G10; C11), including 100726.10+124856.2 (PG~1004+130),
identified as a BAL RLQ and described in Wills et al.~(1999), for
which we here add a previously unpublished {\it HST\/} spectrum to
archival {\it IUE\/} coverage.

\subsection{Radio and optical properties}

The radio and optical properties of the BAL RLQs studied here are
given in Table~1. We use radio and optical/UV monochromatic
luminosities with units of erg~s$^{-1}$~Hz$^{-1}$ at rest-frame 5~GHz
and 2500~\AA, and ${\ell}_{\rm r}$ and ${\ell}_{\rm uv}$ are expressed
as logarithms. Radio-loudness is taken to be the logarithmic ratio of
monochromatic luminosities measured at (rest-frame) 5~GHz and
2500~\AA~(e.g., Stocke et al.~1992; c.f. Kellerman et al.~1998), so
$R^{*}={\ell}_{\rm r}-{\ell}_{\rm uv}$. The optical/UV luminosity is
calculated at rest-frame 2500~\AA~using SDSS photometry [specifically,
the nearest one or two magnitudes to 2500$(1+z)$~\AA, corrected for
Galactic extinction] through comparison to a redshifted composite
quasar spectrum (Vanden Berk et al.~2001) convolved with $ugriz$
filters. The radio luminosity is calculated at rest-frame 5~GHz using
catalog FIRST peak fluxes for the core, and integrated fluxes for the
lobes, assuming typical radio spectral indices of
${\alpha}=-0.3$/${\alpha}=-0.9$ for core/lobe components.

Figure~1 shows \ion{C}{IV} $EW$ versus $R^{*}$ for our sample in the
top panel, along with values for other BAL RLQs (also identified
through matching to FIRST data). A histogram of $EW$ values (from G09)
for BAL RLQs and for BAL RQQs is provided in the lower panel. As
previously noted (see $\S$1), there is a paucity of objects that are
simultaneously strongly radio loud and heavily absorbed. However, our
sample includes objects spanning the full range of $R^{*}$ and $EW$
that are typical of BAL RLQs, and its distribution is representative
of that class.\footnote{Kolmogorov-Smirnov (KS) tests comparing $EW$
give probability $p=0.17$, indicating the distributions are not
inconsistent; similarly, for $R^{*}$ we find $p=0.19$. BAL RLQs and
BAL RQQs have formally inconsistent distributions of $EW$
($p=0.006$).} Figure~2 shows \ion{C}{IV} $v_{\rm max}$ versus
${\ell}_{\rm r}$ for our sample in the top panel, again with values
for other BAL RLQs, with a histogram of $v_{\rm max}$ provided in the
lower panel (from G09; note they truncate BALs beyond 25000
km~s$^{-1}$). Here too our sample spans the full range of parameter
space and is representative of BAL RLQs in general.\footnote{KS tests
comparing $v_{\rm max}$ give probability $p=0.44$, indicating the
distributions are not inconsistent; similarly, for ${\ell}_{\rm r}$ we
find $p=0.16$ when restricting to ${\ell}_{\rm r}>32$.}

Radio maps of the six lobe-dominated BAL RLQs for which we obtained
new HET spectra are provided in Figure~3, from the 1.4~GHz FIRST
survey, with the optical SDSS position indicated with a red cross. The
radio morphology of PG~1004+130 is shown and discussed in
Gopal-Krishna \& Wiita (2000) and Miller et al.~(2006).

\begin{figure}
\includegraphics[scale=0.45]{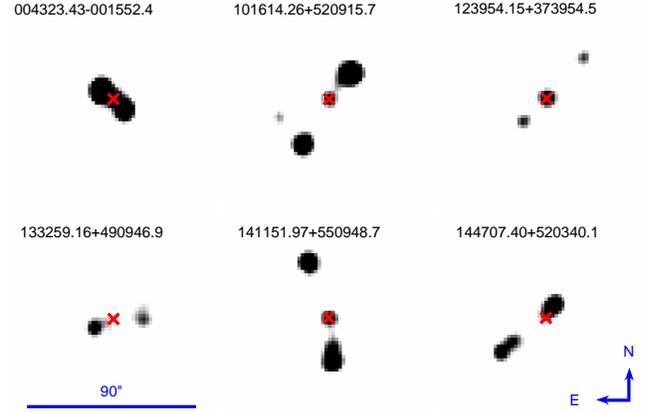}
\caption{\small FIRST 1.4~GHz radio maps of the lobe-dominated BAL
RLQs for which we obtained new HET spectra. The angular resolution is
$\simeq$5$''$. The SDSS position of the quasar core is marked with a
cross. The flux density is plotted on a consistent logarithmic scale.}
\end{figure}

\section{BAL Variability in RLQs}

Our measurements of \ion{C}{IV} absorption equivalent widths for each
absorption trough, in each object, at each epoch, are listed in
Table~2. The associated variability measures (change in equivalent
width, ${\Delta}EW$, and absolute fractional change,
$|{\Delta}EW|/{\langle}EW{\rangle}$, for each pair of spectra are
given in Table~3. Examples of data reduction and analysis are given in
Figures~4, 5, and 6; all spectra are provided in Appendix A. Spectral
variability in a few individual objects of interest is highlighted in
Figures 7, 8, and 9.

\subsection{Observations and data reduction}

The HET observations were obtained in queue-scheduled mode in
2007--2008 and 2011 (see Table~2). The LRS was used for the
spectroscopic observations, in almost all cases with the g2 grating
and typically with a 1.5$''$ slit. This configuration provides a
spectral resolution of $R\simeq$870 (or $\sim$340~km~s$^{-1}$ at
observed 5500\AA), which is sufficiently closely matched to that of
SDSS (typical resolution $R\simeq$1800, or 170~km~s$^{-1}$) to permit
effective evaluation of BAL variability. The HET data were reduced in
standard fashion using the Image Reduction and Analysis Facility
(IRAF\footnote{{\tt http://iraf.noao.edu/iraf/web/}}). In brief, we
generated and subtracted a master bias, using a 30$\times$800 overscan
stripe to account for time-dependent changes; created a normalized
master flat from internal lamp frames, and divided this into the
object frames to remove pixel-to-pixel sensitivity variation;
subtracted the sky background using apertures of 100 pixel width
offset by 50 pixels from the object (shifted if necessary to avoid
background objects); manually identified and interpolated over cosmic
rays; traced the aperture and extracted a one-dimensional spectrum;
applied wavelength calibration based on ThAr comparison spectra;
shifted wavelengths to the heliocentric frame; applied relative
flux-calibration based on a standard star (absolute flux calibration
is not required, as we use normalized spectra --- this step is
primarily useful to remove the blaze function).

The SDSS spectra were downloaded\footnote{{\tt http://www.sdss.org}}
and wavelengths were converted to air values for direct comparison to
our HET spectra. For both HET and SDSS or other archival data, we
modeled the continuum with a low-order (second or third) Chebyshev
polynomial, masking the L$\alpha$ and \ion{N}{V}, \ion{Si}{IV}, and
\ion{C}{IV} emission line regions as well as any obvious BAL
absorption, and additionally imposing strict multi-pass $\sigma$
rejection criteria to bypass less prominent lines or weak BALs. The
resulting best-fit continuum polynomial was then divided out to
normalize the spectra near the \ion{C}{IV} region. This implicitly
dereddens the spectra. Typically the same polynomial constraints
(wavelength ranges for fitting, polynomial order) were used to
normalize all spectra for a given object.

\begin{figure}
\includegraphics[scale=0.45]{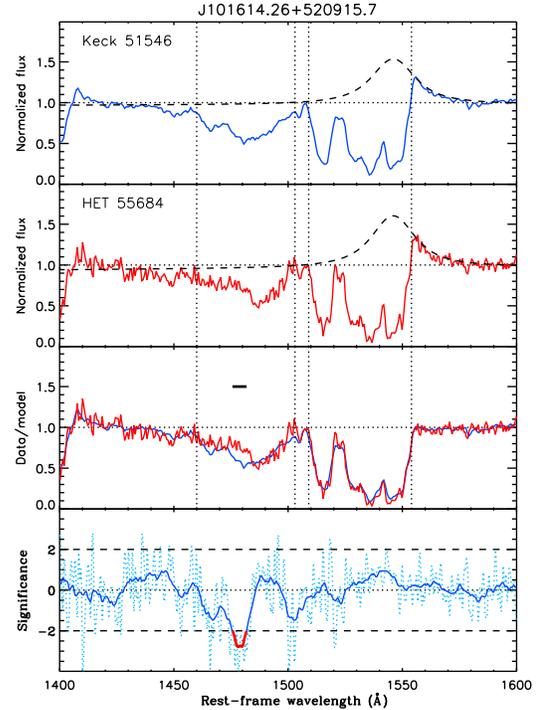} 
\caption{\small Illustration of measurement of BAL properties. The
  \ion{C}{IV} emission line is modeled with a Voigt profile (dashed
  lines, top two panels) and divided out, then the equivalent width is
  calculated within the BAL region (vertical dotted lines). In this
  case the BAL varied significantly at 1476--1482~\AA~(bottom panel)
  between the two observations. Variability in the bottom panel is
  plotted in units of $\sigma$ (dotted cyan line) as the difference
  between the normalized spectra divided by the typical uncertainty
  (as measured empirically in a non-varying flat region of the
  continuum), then smoothed (solid blue line) prior to checking for
  any $>2\sigma$ changes (solid red line overplotted, where present).}
\end{figure}

\begin{figure}
\includegraphics[scale=0.45]{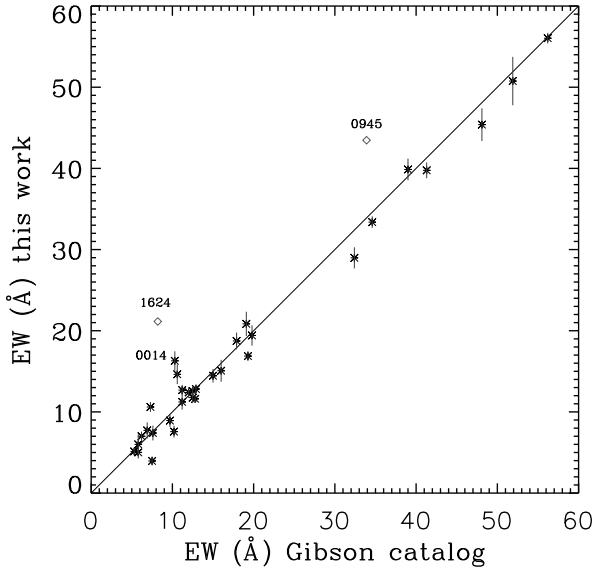}
\caption{\small Our equivalent width measurements for BAL RLQs
  compared with those determined by G09 for the same SDSS
  quasars. There is close agreement; the mean difference is
  0.12~\AA. The open diamonds mark objects for which the full BAL
  extends beyond the 25000 km~s$^{-1}$ boundary considered in
  G09. Typical 1$\sigma$ statistical errors on the G09 measurements
  are $\pm1$~\AA.}
\end{figure}

\subsection{Measurement of absorption properties}

Wavelengths are converted to the rest-frame of the quasar using the
improved redshifts calculated from SDSS spectra by Hewett \& Wild
(2010) where available, or else from other sources including the
NASA/IPAC Extragalactic Database.\footnote{{\tt
    http://ned.ipac.caltech.edu/}} We restrict our focus to BAL
variability, and so any emission-line variability or remaining
residual continuum variability is modeled out. We fit a Voigt profile
to the \ion{C}{IV} emission line in each spectrum (including a linear
residual continuum term), choosing this model for convenience in
matching observations but ascribing no particular physical
significance to the functional form (see also, e.g., G10). The
emission-line parameters are calculated for the highest
signal-to-noise observation and then applied to the other
observations; for a few cases in which the emission profile has
obviously varied, the lower signal-to-noise observations are fit
independently. The continuum placement and slope is always permitted
to vary. In cases for which the BAL absorption extends within the
emission-line region, it often proved necessary to adjust manually the
fit parameters to obtain a satisfactory model. Similar fitting to an
empirical RLQ spectral template from Kimball et al.~(2011) provided
the initial parameters for the \ion{C}{IV} emission line profile, and
was adhered to as closely as possible where absorption prevented
automated fitting. To simplify analysis of the absorption properties,
the \ion{C}{IV} emission-line model is then divided out, and BAL
properties are measured from the resulting ratio spectrum.

The edges of a BAL are here defined to be the wavelengths at which the
BAL intersects 90\% of the normalized continuum, following Weymann et
al.~(1991). An example is show in Figure~4. The only exception occurs
for J123411.74$+$615832.5, for which the HET spectrum does not fully
cover the BAL; here the maximum considered velocity is fixed to the
limit of high S/N HET coverage. BALs are required to have velocity
widths $\simgt$2000~km~s$^{-1}$, but are allowed to be present at
velocities ranging from zero (or even slightly redshifted) to
30000~km~s$^{-1}$, beyond the classical 3000 to 25000~km~s$^{-1}$
boundaries often imposed (e.g., Weymann et al.~1991; G09). The
equivalent width of each \ion{C}{IV} BAL is then calculated as a sum
over all pixels within the BAL region (recall all artifacts have been
removed). The error on the equivalent width is calculated including
contributions from three sources: fitting error in the continuum
placement, statistical noise in the spectrum (which may be higher
within the BAL region), and measurement uncertainty in the BAL edge
determination. The first term is estimated as the standard deviation
of the data within a flat 50~\AA~region, and is applied
pixel-by-pixel. The next term is estimated as the deviation from a
smoothed spectrum at each point; it is added in quadrature to the
first term. The final term is estimated by recalculating the
equivalent width using BAL edges shifted by $\pm$0.5~\AA; the absolute
value of the difference is added in quadrature to the sum of the first
and second terms. Our equivalent width measurements agree well with
those conducted by Gibson et al.~(2009b) for the SDSS quasars (at a
given epoch) in common (Figure~5). After removing two objects with
BALs extending beyond the 25000 km~s$^{-1}$ boundary considered by
Gibson et al.~(2009b), the mean difference is only 0.12\AA. The
largest remaining outlier is an object for which G09 exclude a
low-velocity component which our continuum/emission fit includes in
the primary BAL.

The time interval between spectra in the quasar rest frame is
determined as $\tau=(MJD_{\rm new}-MJD_{\rm old})/(1+z)$, where the
modified Julian dates correspond to the midpoints of the respective
observations. The average equivalent width for a given object is
simply ${\langle}EW{\rangle}=0.5(EW_{\rm new}+EW_{\rm old})$, and the
change in equivalent width is ${\Delta}EW=EW_{\rm new}-EW_{\rm old}$,
so that a positive value of ${\Delta}EW$ signifies an increase in $EW$
(e.g., a BAL becoming deeper). The absolute fractional change in
equivalent width is $|{\Delta}EW|/{\langle}EW{\rangle}$.

Potential velocity shifts within the BAL structure were assessed
through shifting one spectrum relative to the other and minimizing the
squared residual (Hall et al.~2007). An example is shown in
Figure~6. No significant examples of bulk velocity shifts in BAL
features were found, with limits typically around
$\simlt100$~km~s$^{-1}$. For the longest rest-frame timescales within
our sample of $\sim$1000~d, this limits average acceleration to
$\simlt$0.1~cm~s$^{-2}$.

\begin{figure}
\includegraphics[scale=0.45]{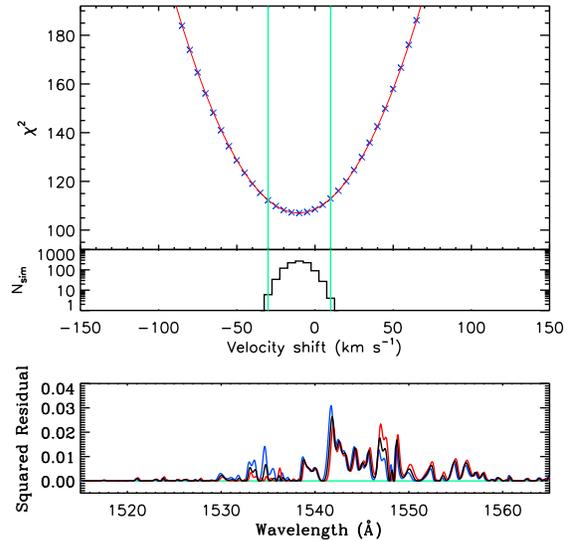}
\caption{\small Illustration of assessment of velocity shifts between
  epochs. One spectrum is shifted relative to the other to identify
  the optimal wavelength offset (lowest squared residual; bottom); in
  this case, the 99\% confidence region includes zero
  (top). Simulations confirm that the errors are consistent (middle).}
\end{figure}

\subsection{Notes on individual objects}

Here we describe individual objects of interest within our sample of
BAL RLQs. For additional notes on some objects, see also $\S$3.3 of
Miller et al.~(2009).

{\it Known LoBAL quasars\/}: Four of these BAL RLQs have absorption at
ions other than \ion{C}{IV} or \ion{Si}{IV} in the SDSS catalog of
G09: J083925.61+045420.2 and J132139.86$-$004151.9 have \ion{Al}{III}
absorption and lack SDSS \ion{Mg}{II} coverage, while
J094513.89+505521.8 and J123954.15+373954.5 have both \ion{Al}{III}
and \ion{Mg}{II} absorption. J141546.24+112943.4 is in DR6+ but not
DR5 so it is not included in G09, but inspection of the SDSS spectrum
shows \ion{Al}{III} absorption, previously identified as broad by
Hazard et al.~(1984). Note that for the BAL RLQs with $z>2.1$ the
\ion{Mg}{II} line is redshifted beyond SDSS coverage.

{\it Mini-BAL absorption\/}: In addition to its high-velocity BAL,
J162453.47$+$375806.6 also has lower-velocity absorption described by
Benn et al.~(2005) as a mini-BAL. Some of the narrower BALs studied
here are elsewhere (e.g., Rodriguez Hidalgo et al.~2012) considered as
mini-BALs, including J115944.82$+$011206.9.

{\it Lobe-dominated RLQs\/}: Seven of the BAL RLQs studied here have
radio emission dominated by bright radio lobes
(Figure~3). J004323.43$-$001552.4 (Gregg et al.~2006; Brotherton et
al.~2006, 2011), J133259.16$+$490946.9, and J144707.40$+$520340.1
(Gregg et al.~2006) are all included in the catalog of SDSS FR~II
quasars of de~Vries et al.~(2006), while J100726.10+124856.2 (Wills et
al.~1999) and J101614.26+520915.7 (Gregg et al.~2000) were among the
first-identified BAL RLQs. J123954.15+373954.5 and J141151.97+550948.7
are also lobe-dominated. These objects are presumably viewed at
relatively large inclinations (e.g., Wills \& Brotherton~1995).

{\it Small-scale structure\/}: Several of the core-dominated RLQs
display one-sided jet emission on mas scales, including
J115944.82$+$011206.9 (here a short counterjet feature is also present
near the core; Montenegro-Montes et al.~2009), J141334.38$+$421201.7
(Liu et al.~2008), and J162453.47$+$375806.6 (Benn et al.~2005,
Montenegro-Montes et al.~2009).

{\it Flat radio spectrum\/}: Several RLQs have 5~GHz Green Bank data
(or low-frequency observations) that suggest they have flat
(${\alpha}\simgt-0.3$) radio spectra, including J074610.50$+$230710.7,
J085641.56$+$424253.9, J105416.51$+$512326.0, J133701.39$-$024630.2,
and J141334.38$+$421201.7.  A flat radio spectrum in a core-dominated
RLQ indicates a relatively low inclination. The radio spectrum of
J115944.82$+$011206.9 is double-peaked (Montenegro-Montes et al.~2009)
so this may be a younger source.

\begin{figure}
\includegraphics[scale=0.42]{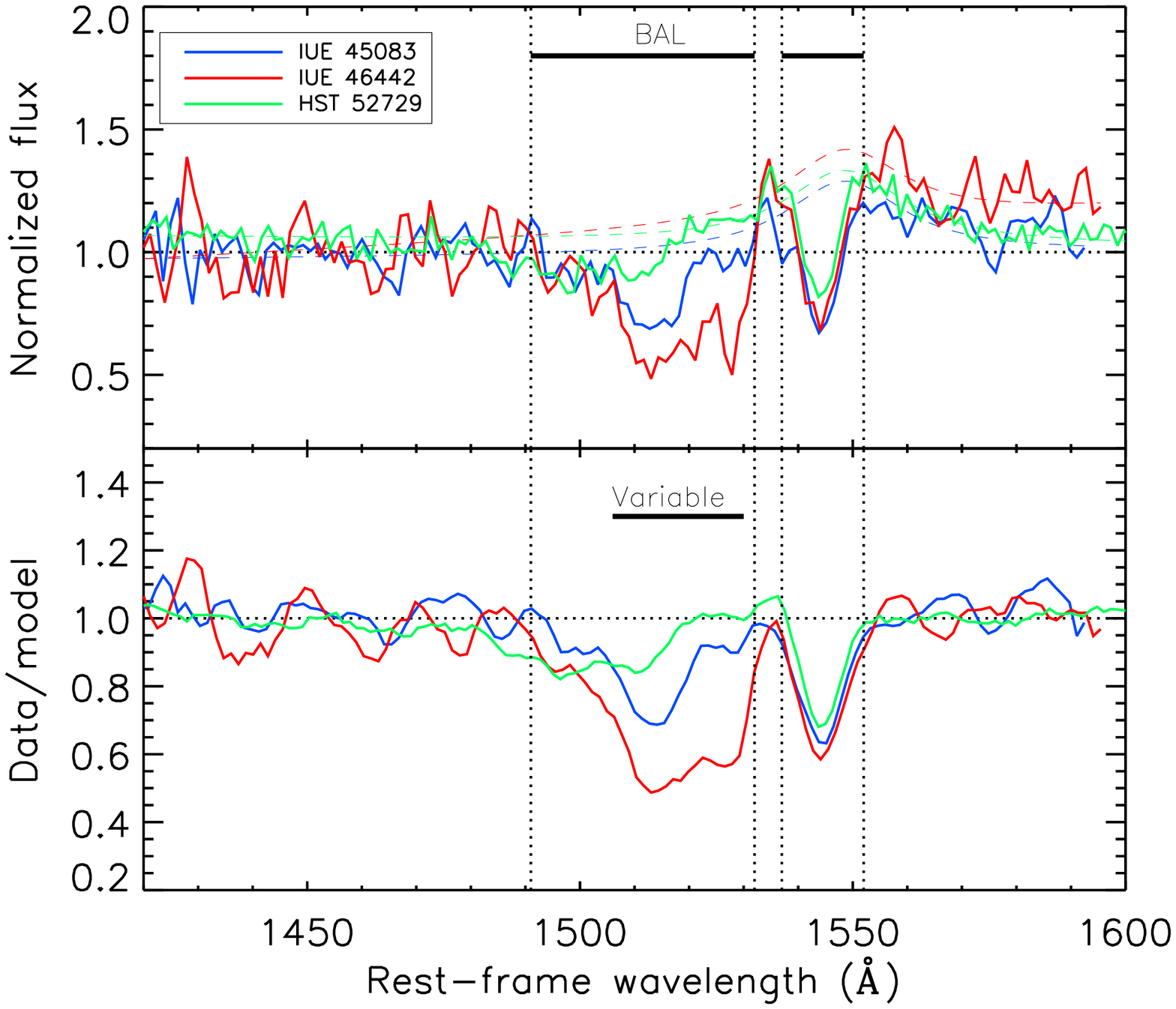}
\includegraphics[scale=0.40]{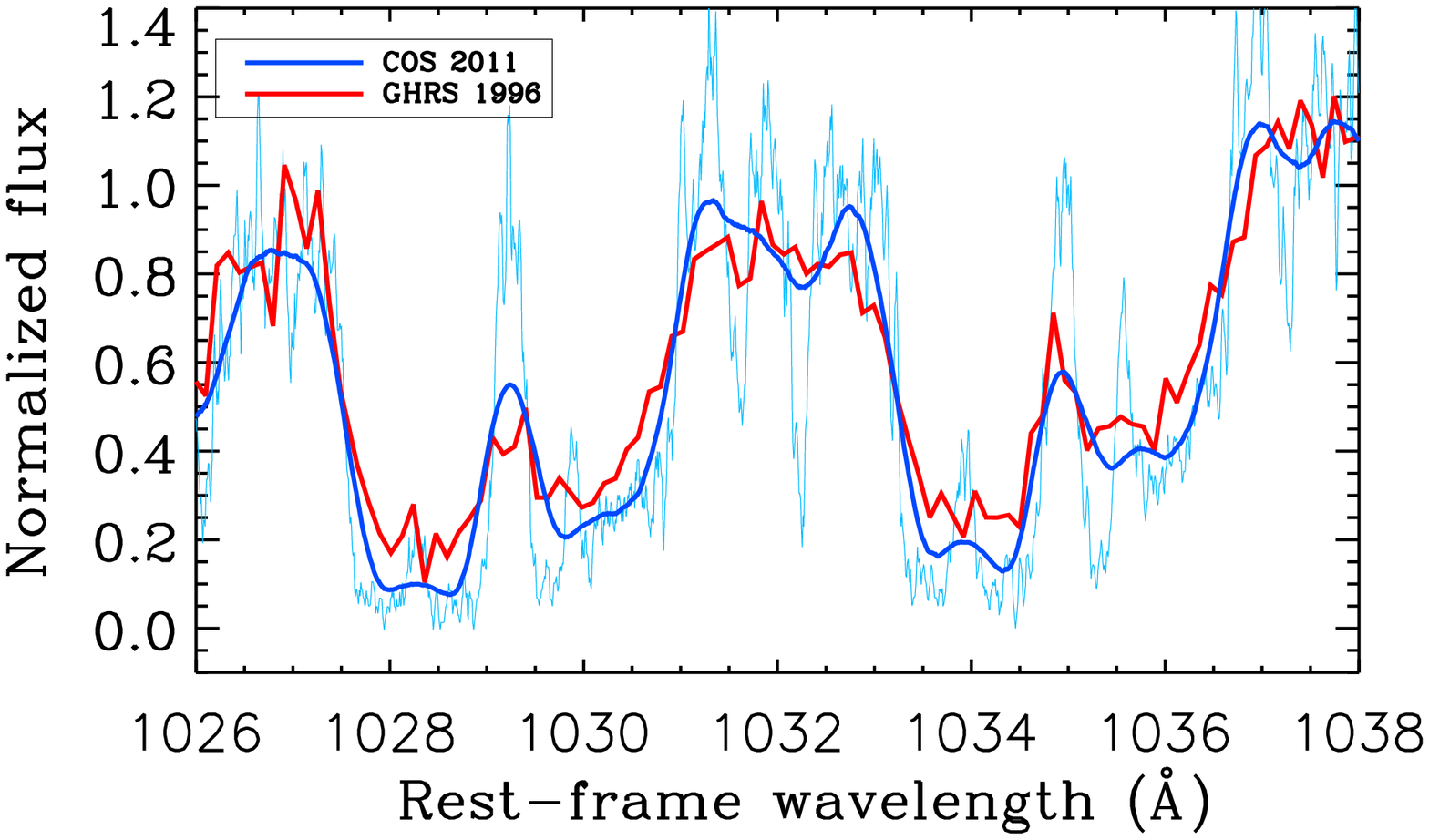} 
\caption{\small PG~1004+130 BAL absorption variability. Top, upper:
  observed \ion{C}{IV} spectrum with continuum and emission line model
  shown. Top, lower: normalized spectrum boxcar smoothed by 5 pixels
  (6.8~\AA). Note the increase and subsequent decrease in absorption
  depth within a segment of the high-velocity outflow component over
  16.9 rest-frame years. Bottom: O~VI low-velocity absorption with COS
  (cyan) and GHRS (red); when the COS resolution is degraded (blue)
  with a Gaussian kernal to match the GHRS spectrum, only slight
  changes in the depths of the troughs are found, in contrast to the
  dramatic \ion{C}{IV} variability.}
\end{figure}

\begin{figure}
\includegraphics[scale=0.45]{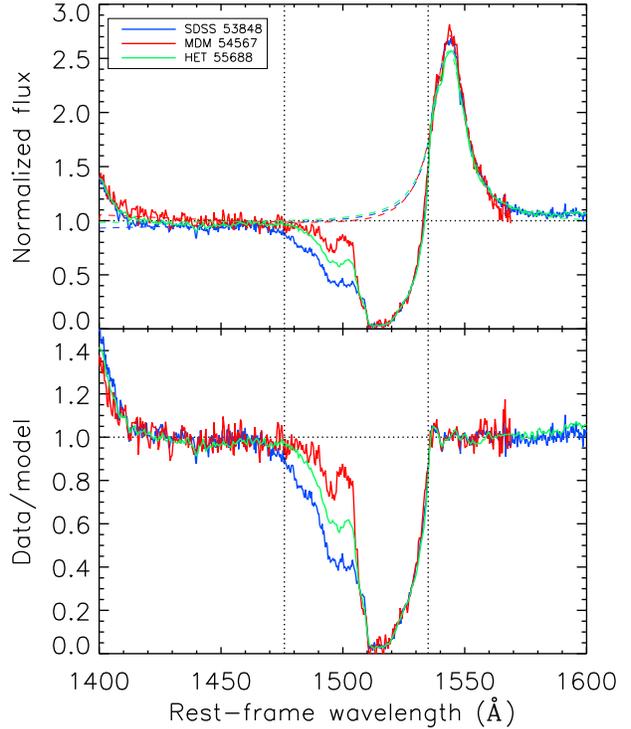} 
\caption{\small J141546.24+112943.4 spectra from SDSS, MDM, and
HET, showing a strong decrease in absorption strength followed by a
partial recovery toward increased absorption. See also C11.}
\end{figure}

\begin{figure*}
\includegraphics[scale=0.85]{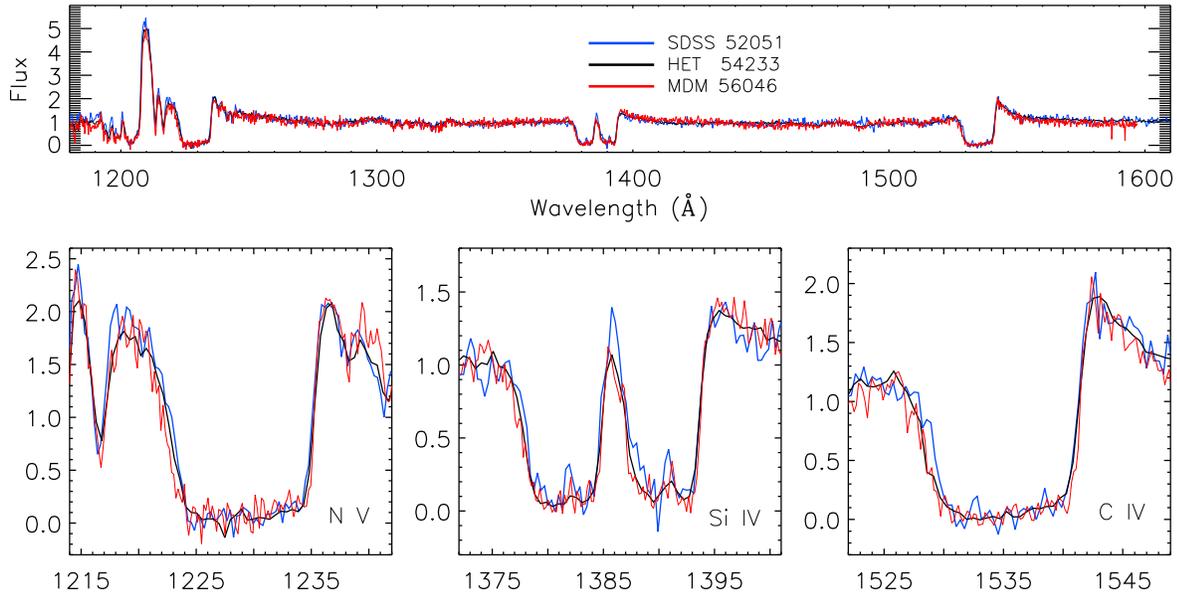} \caption{SDSS, HET, and MDM
  spectra of the BAL RLQ J160943.35+522550.8 (top), with details of
  the \ion{C}{IV}, \ion{Si}{IV}, and \ion{N}{V} absorption regions
  (bottom). The BALs in the HET spectrum (black) extend to slightly
  greater velocities than in the earlier SDSS spectrum (blue), but
  align with the later MDM spectrum (red). This illustates that an
  apparent acceleration between two epochs might instead result from a
  change in depth within a high-velocity segment, here identifiable in
  the third epoch data.}
\end{figure*}

{\it J100726.10$+$124856.2\/}: This low-redshift lobe-dominated BAL
RLQ (PG~1004+130, $z=0.241$, $M_{\rm B}=-25.6$; Wills et al.~1999;
Brandt et al.~2000) is known to show variable \hbox{X-ray} absorption
(Miller et al.~2006). Here we show that it also possesses strongly
variable \ion{C}{IV} absorption (Figure~7), for example with the
primary BAL greatly diminishing in depth from the January 1986 {\it
  IUE\/} observation to the March 2003 {\it HST\/}/STIS spectrum
($\sim$14 rest-frame years). PG~1004+130 displays the greatest
absolute and fractional equivalent-width variations in our survey of
BAL RLQs. The lower-velocity absorption is also seen at higher
ionizations, including \ion{O}{VI}. A comparison of the \ion{O}{VI}
absorption from recent FUV COS/G130M observations with a GHRS spectrum
from 1996 shows perhaps a slight increase in the depths of the troughs
(Figure~7), but this modest variability is of substantially lower
amplitude than is seen in the primary \ion{C}{IV} BAL at higher
velocities. The slower interior flow is more exposed to direct
high-energy radiation\footnote{In BAL RLQs, this may include a
  small-scale jet contribution (e.g., Miller et al.~2009).} and is
consequently more highly ionized; the \ion{O}{VI} doublets are
optically thick and partially covering (Wills et al.~1999).

{\it 141546.24+112943.4\/}: This is the extensively studied
``cloverleaf'' quasar H1413$+$117, a gravitationally lensed object
that is split into 4 images separated by $\sim$1$''$. The A image
shows enhanced X-ray emission, suggestive of microlensing (Chartas et
al.~2004), and based on CO emission the quasar is surrounded by a
rotating molecular disk (Venturini \& Solomon~2003). This modestly
radio-loud ($R^{*}=1.09$) quasar displays variability within the
higher-velocity half of the primary BAL while the lower-velocity
segment remains constant. Between the SDSS and subsequent MDM spectra
(MDM data from C11, see that work for details), the absorption
decreases in strength, but by the latest HET spectrum the absorption
is again increasing in strength (Figure~8). As noted above, this is
one of the few known LoBALs in our sample.

{\it J160943.35+522550.8\/}: This object showed an apparent increase
in the maximum velocity of the BAL outflow between the SDSS and HET
observations. While the red edge remained fixed, the blue edge extends
to shorter wavelengths in the later HET data, in not only \ion{C}{IV}
but also \ion{Si}{IV} and \ion{N}{V} (Figure~9). Systematic changes in
the velocity structure of a BAL are unusual, for either BAL RQQs or
BAL RLQs (e.g., Hall et al.~2007; Gibson et al.~2008). We obtained a
new MDM spectrum to check whether a high-velocity component had
accelerated between epochs, leaving the main absorption trough
unchanged due to the putative presence of additional absorbers
partially overlapping in velocity space and together fully covering
the continuum. The MDM observations were carried out using the 2.4m
Hiltner telescope with the Ohio State Multi-Object Spectrograph (VPH
grism, center 1.2$''$ slit). The MDM spectrum aligns with the HET
spectrum (Figure~9), arguing against acceleration and suggesting
instead that a high-velocity component was always present but
increased in strength (e.g., through a highly ionized clump cooling to
a greater \ion{C}{IV} opacity), or that a high-velocity cloud
transversely entered the line of sight. This illustrates the
importance of obtaining a third epoch spectrum to check cases of
potential acceleration.

\begin{figure*}
\includegraphics[scale=0.75]{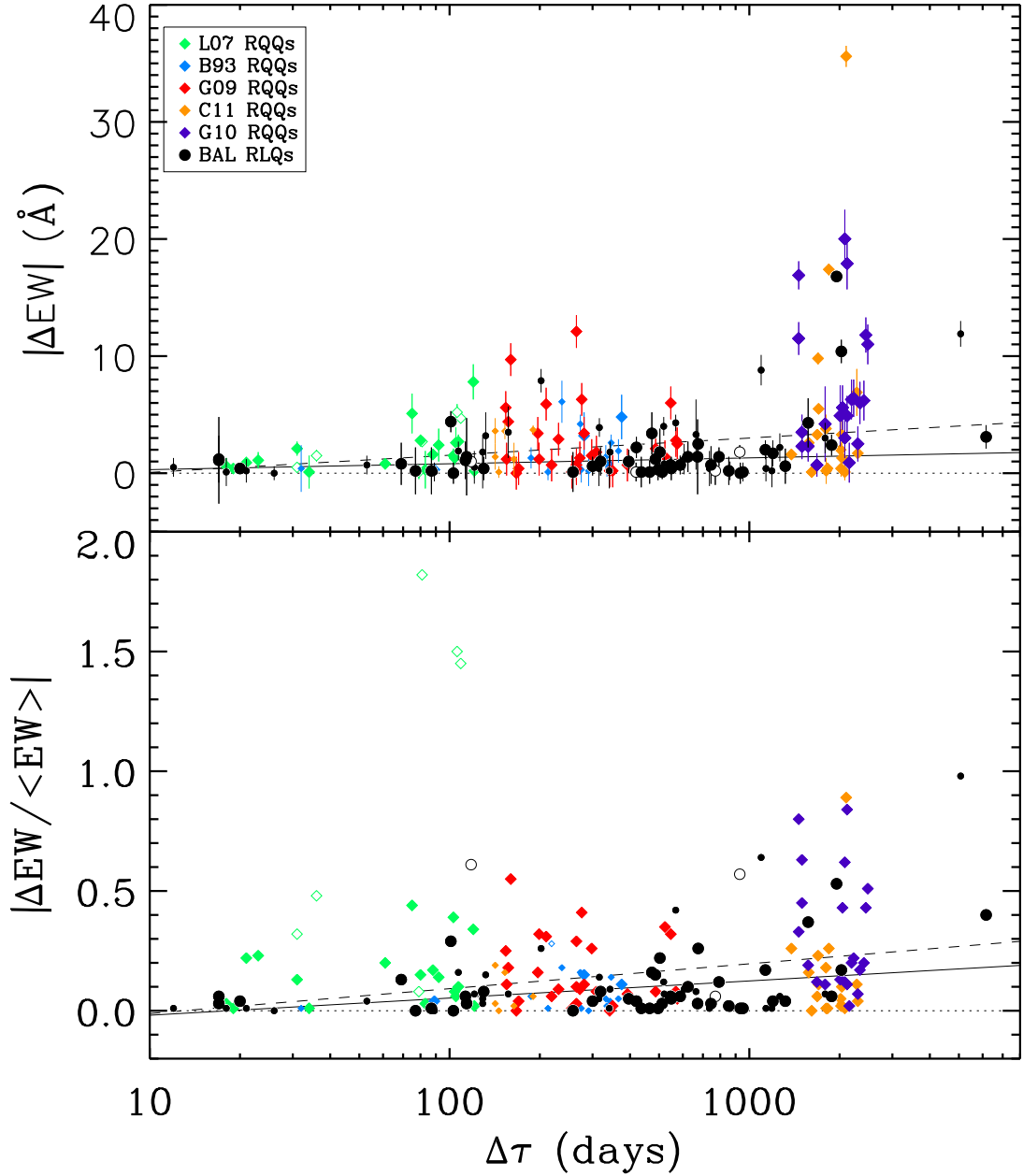}
\caption{\small Change in BAL equivalent width and absolute fractional
  change in BAL equivalent width versus interval between spectroscopic
  observations for RQQs (diamonds) and RLQs (circles). RQQ subsamples
  plotted in green/blue/red/orange/purple are from
  L07/B93/G09/C11/G10, respectively, and open symbols have
  $EW<3.5$~\AA. For objects with multiple pairs the shorter-separation
  measurements are plotted using smaller markers. The dashed and solid
  lines show linear fits to RQQs and RLQs; the dotted line is zero.}
\end{figure*}

\section{Analysis and Discussion}

Below, we investigate the distributions of absolute change in
equivalent width $|{\Delta}EW|$ and absolute fractional change in
equivalent width $|{\Delta}EW|/{\langle}EW{\rangle}$ as a function of
rest-frame interval between spectral observations (${\Delta}{\tau}$;
Figure~10) and of average BAL equivalent width
(${\langle}EW{\rangle}$; Figure~11). Variability is also assessed with
respect to velocity width (Figure~12), and comparisons are made to BAL
variability patterns within RQQs. Optical continuum variability is
assessed and quantified for BAL RLQs (Figure~13) and BAL RQQs
(Appendix~B, Figures 1 and 2) and compared across classes (Figures 14
and 15). For RLQs, the impact of radio loudness or luminosity ($R^{*}$
or ${\ell}_{\rm r}$) upon BAL variability is investigated
(Figure~16). The longest-separation total BAL absorption properties
for BAL RLQs and for the comparison sample of BAL RQQs are given in
Tables~1 and 4, respectively.

For a portion of the analysis it is convenient to distinguish between
subsamples of BAL RLQs grouped by radio properties. RLQs are separated
into core-dominated or lobe-dominated, low or high radio luminosity
(at ${\ell}_{\rm r}=33$), and low or high radio loudness (at
$R^{*}=2$). Note that while the archival coverage of PG~1004+130
(lobe-dominated, low radio luminosity, high radio loudness) extends to
longer timescales than are typical within our sample, for this object
the variability over $\sim$1000~d is actually larger than for the
longest separation $\sim$6000~d measurement used.

Median and mean properties, along with Kolmogorov-Smirnov (KS) test
probabilities for selected comparisons, are provided in
Table~5. Correlation likelihoods (non-parametric Kendall $\tau$ and
Spearman $\rho$) and coefficients along with best-fit linear
regression slopes and errors (calculated using the {\tt IDL}
robust\_linefit routine) for each tested relationship are listed in
Table~6.

\begin{figure*}
\includegraphics[scale=0.73]{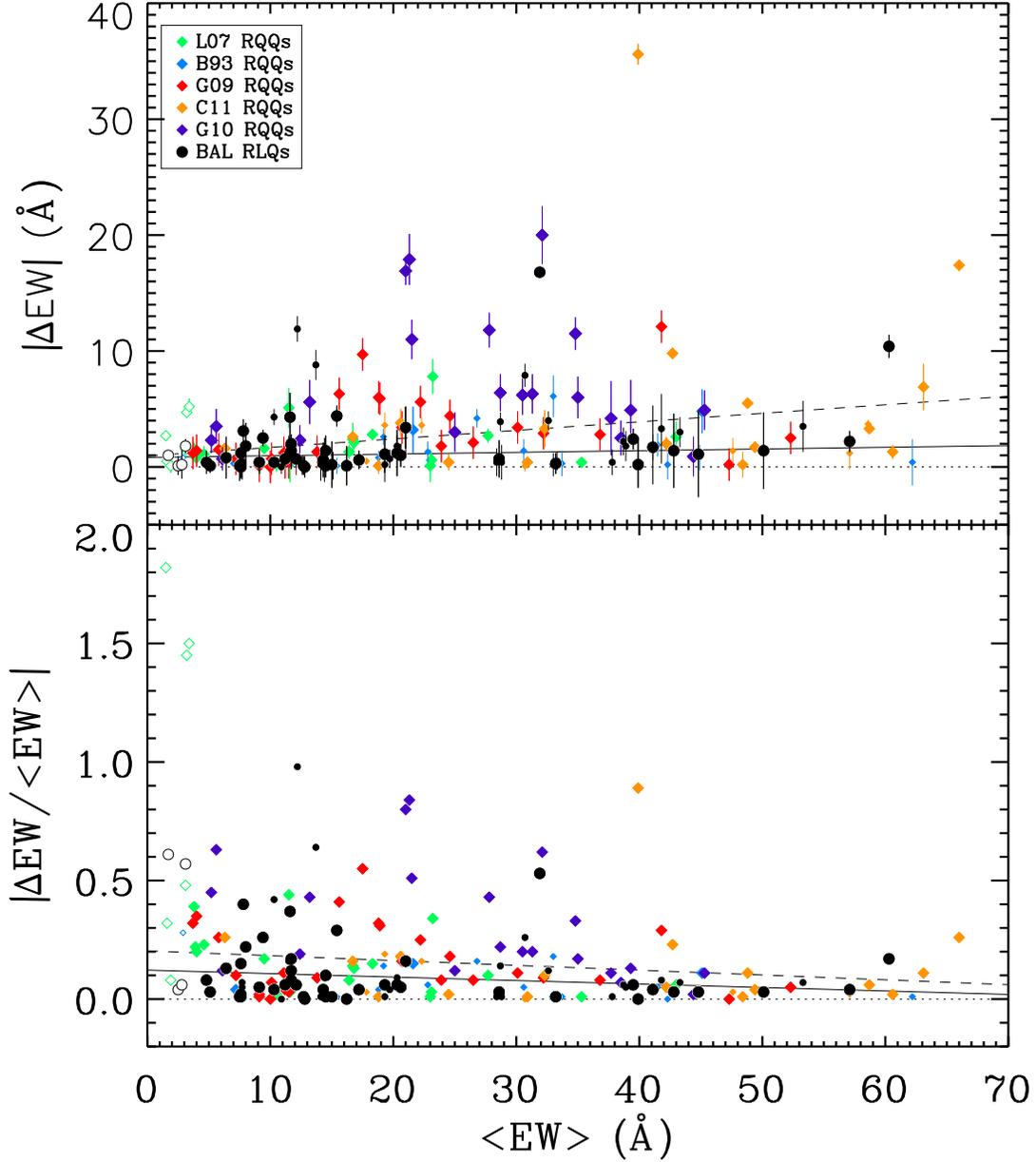} 
\caption{\small Change in BAL equivalent width versus average BAL
equivalent width for RQQs (diamonds) and RLQs (circles). Symbols as in
Figure~10. There is a tendency ($\simgt$95\% confidence) toward greater
$|{\Delta}EW|$ within stronger BALs, for both RLQs and RQQs.}
\end{figure*}

\subsection{Comparison to BAL RQQs}

To avoid potential biases arising from repeated sampling of particular
objects (recall we have 78 pairs of spectra of 46 BAL RLQs), the
longest-separation measurement of variability available for each of
the 46 BAL RLQs is used for all statistical comparisons to RQQs. We
constructed a comparison sample of BAL RQQs from previous studies of
BAL variability (B93, L07, G09, G10, and C11), for verified
radio-quiet\footnote{We checked radio-loudness against FIRST data
  where possible, otherwise against the NASA/IPAC Extragalactic
  Database (NED; {\tt http://ned.ipac.caltech.edu/}); note that the
  RQQs LBQS 0055+0025 and [HB89] 2225$-$055 are near unassociated
  radio sources.} quasars; recall from $\S$2.1 that the handful of
RLQs covered in these studies are included in our radio-loud
sample. In particular, we use 25 pairs of spectra from dual-epoch SDSS
measurements from L07 and another 28 from G09 (requiring
${\Delta}\tau>150$~d, hence distinct from L07); 16 pairs of spectra
from dual-epoch Lick measurements from B93 (some additional
measurements from this reference are superceded by C11); 21 pairs of
spectra from Lick/SDSS/HET spectra from G10; and 25 pairs of spectra
from Lick/SDSS/MDM spectra from C11. Together, these studies cover
short (20--120~d; L07), short/intermediate (80--400~d; B93),
intermediate (130--600~d; G09), intermediate/long (100-200~d and
1300--3000~d; C11), and long (1300--2500~d; G10) timescales. The
combined sample of RQQs then includes 115 pairs of BAL absorption
measuments. From these, we construct a longest-separation sample with
a single measurement of BAL variability for each of the 94 unique BAL
RQQs.

Four RLQs and six RQQs have small absorption equivalent widths
${\langle}EW{\rangle}<3.5$~\AA~which are not typical of BALs;
following Gibson et al.~(2008), we compare RQQs and RLQs after removal
of these objects. The resulting filtered longest-separation samples of
42 RLQs and 88 RQQs with ${\langle}EW{\rangle}\ge3.5$~\AA~span similar
ranges in redshift and luminosity, but have inconsistent distributions
(KS test $p<0.03$ of being drawn from the same underlying population)
of both ${\Delta}\tau$ and ${\langle}EW{\rangle}$, in the sense that
these RQQs cover longer timescales and have larger BAL equivalent
widths. We constructed a matched group of 42 RQQs through selecting
objects with ${\Delta}\tau$ and ${\langle}EW{\rangle}$ values similar
to those of the filtered RLQs, without consideration of any
variability properties (the KS probabilities for the matched RQQ
sample are now $p=0.26$ and $p=0.56$ for ${\Delta}\tau$ and
${\langle}EW{\rangle}$, respectively). The filtered samples of 42 RLQs
and 88 RQQs were also divided into groups of short and long timescale
(at ${\Delta}\tau=500$~d, which is approximately the median timescale
for both samples) and moderate and large average equivalent width (at
${\langle}EW{\rangle}=20$~\AA, which is approximately the median
equivalent width for the BAL RQQs).

In general, BAL variability within RLQs appears similar to that within
RQQs. Qualitatively, variability within RLQs, when observable,
typically consists of a modest change in the absorption depth, often
within a discrete section of the full trough (Figures 4, 7, 8, and
Appendix~A). Velocity shifts in the structure of BALs appear to be
rare (one candidate from our 46 BAL RLQs; Figure~9). These are similar
to established tendencies within BAL RQQs ($\S$1). Quantitatively,
prior to filtering or matching, the absolute change in $EW$ or
fractional variability is lower within RLQs (e.g., the mean
$|{\Delta}EW|/{\langle}EW{\rangle}$ is 0.12$\pm$0.02 for RLQs versus
0.24$\pm$0.03 for RQQs). After filtering out objects with
${\langle}EW{\rangle}<3.5$~\AA, the fractional variability in RLQs is
still smaller (0.10$\pm$0.02 versus 0.19$\pm$0.02), and this
difference persists in the matched RQQs (0.17$\pm$0.03; this is a
$\sim$2$\sigma$ difference); the KS test probability of $p=0.01$ is
likewise marginally inconsistent with RLQs and matched RQQs possessing
similar BAL variability. The percentage of RLQs displaying significant
BAL variability is 21\%$\pm$7\% (Poisson errors; Table~3 and
Appendix~A), lower than is typical for BAL RQQs on similar timescales
(e.g., C11).

It is possible that this comparison could be influenced by systematic
differences in how BAL variability is measured across different
studies; for example, our approach of locking continuum and emission
line fit parameters as well as BAL edges between epochs wherever
possible may produce lower changes in $EW$ than would result from
completely independent fitting and measurement at each epoch. An
additional point of concern is that our identification of variability
is sensitive to noise. There is an apparent anti-correlation between
BAL variability and optical magnitude (at $\sim2\sigma$ significance)
for the combined RLQ and RQQ sample. However, the optical magnitudes
of the RLQs are similar to those of the matched RQQs (KS test
probability $p=0.54$), with means of $18.46\pm0.14$ and
$18.34\pm0.10$, respectively. We conservatively interpret our results
to indicate that BAL RLQs show similar or perhaps decreased BAL
variability as compared to BAL RQQs. This is consistent with the
findings of Filiz Ak et al.~(2013) of no significant differences in
variability, for a smaller sample of BAL RLQs.

BALs in RLQs are more likely to vary and display a greater variability
amplitude on longer timescales (Figure~10), similar to established
trends for BAL RQQs ($\S$1). The mean
$|{\Delta}EW|/{\langle}EW{\rangle}$ is $0.13\pm0.03$ ($0.06\pm0.02$)
for ${\Delta}\tau\ge500$~d ($<500$~d). The corresponding mean values
for a matched sample of BAL RQQs are somewhat greater ($0.23\pm0.04$
and $0.15\pm0.02$, respectively), although within the longer timescale
bin the full distributions of $|{\Delta}EW|/{\langle}EW{\rangle}$ are
not inconsistent. Kendall and Spearman tests also find a significant
if only moderately strong correlation between ${\Delta}\tau$ and
$|{\Delta}EW|$ (Table~6) for both RLQs and RQQs. The best-fit slope
for $|{\Delta}EW|$ as a function of $\log{{\Delta}\tau}$ is greater
for RQQs. Both BAL RLQs and BAL RQQs also tend to have larger absolute
(but not fractional) changes in equivalent width within stronger BALs
(Figure~11). The mean ${\Delta}EW$ for RLQs is
3.2$\pm$1.2~\AA~(1.1$\pm$0.2~\AA) for
${\langle}EW{\rangle}\ge20$~\AA~($<20$~\AA). Correlation tests again
provide agreement in identifying a significant if moderate trend for
both RLQs and RQQs, again with larger best-fit slopes for RQQs. Note
that the mean and median timescales are similar in the two groupings
of RLQs split by average BAL equivalent width, so we can be confident
that this is a distinct trend from the correlation with timescale (the
average equivalent widths are also similar between the RLQ groups
split by timescale). This is not the case for the RQQs, so here the
RLQs provide a cleaner demonstration of the trends (previously
discovered in RQQs; see $\S$1) toward increasing BAL variability on
longer timescales or (in an absolute but not fractional sense) within
stronger BALs.

\begin{figure}
\includegraphics[scale=0.38]{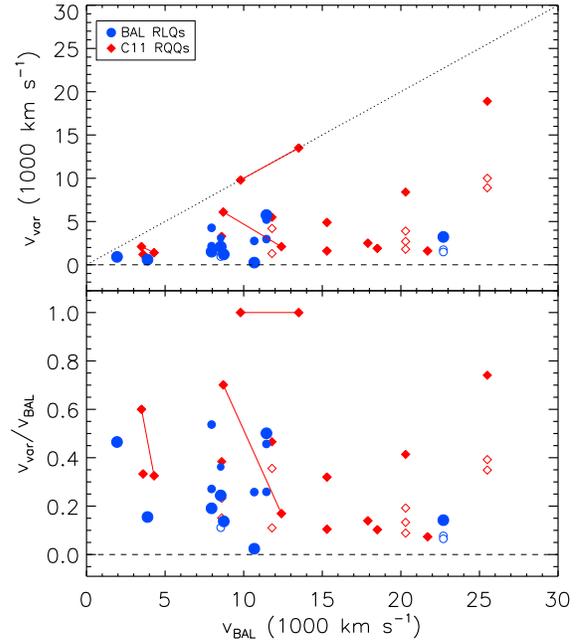} 
\caption{\small Illustration of velocity width over which BALs varied
  for RLQs (blue circles) and RQQs from C11 (red diamonds). The
  open symbols are the sub-sections of a given trough that varied. For
  the RQQs, 3 objects have variability within two distinct troughs
  (connected with lines). For RQQs we only plot the longest separation
  measurements for clarity. For the RLQs, the longest separation
  measurement is plotted as larger symbols, the shorter epoch(s) as
  smaller symbols.}
\end{figure}

\begin{figure*}
\includegraphics[scale=0.75]{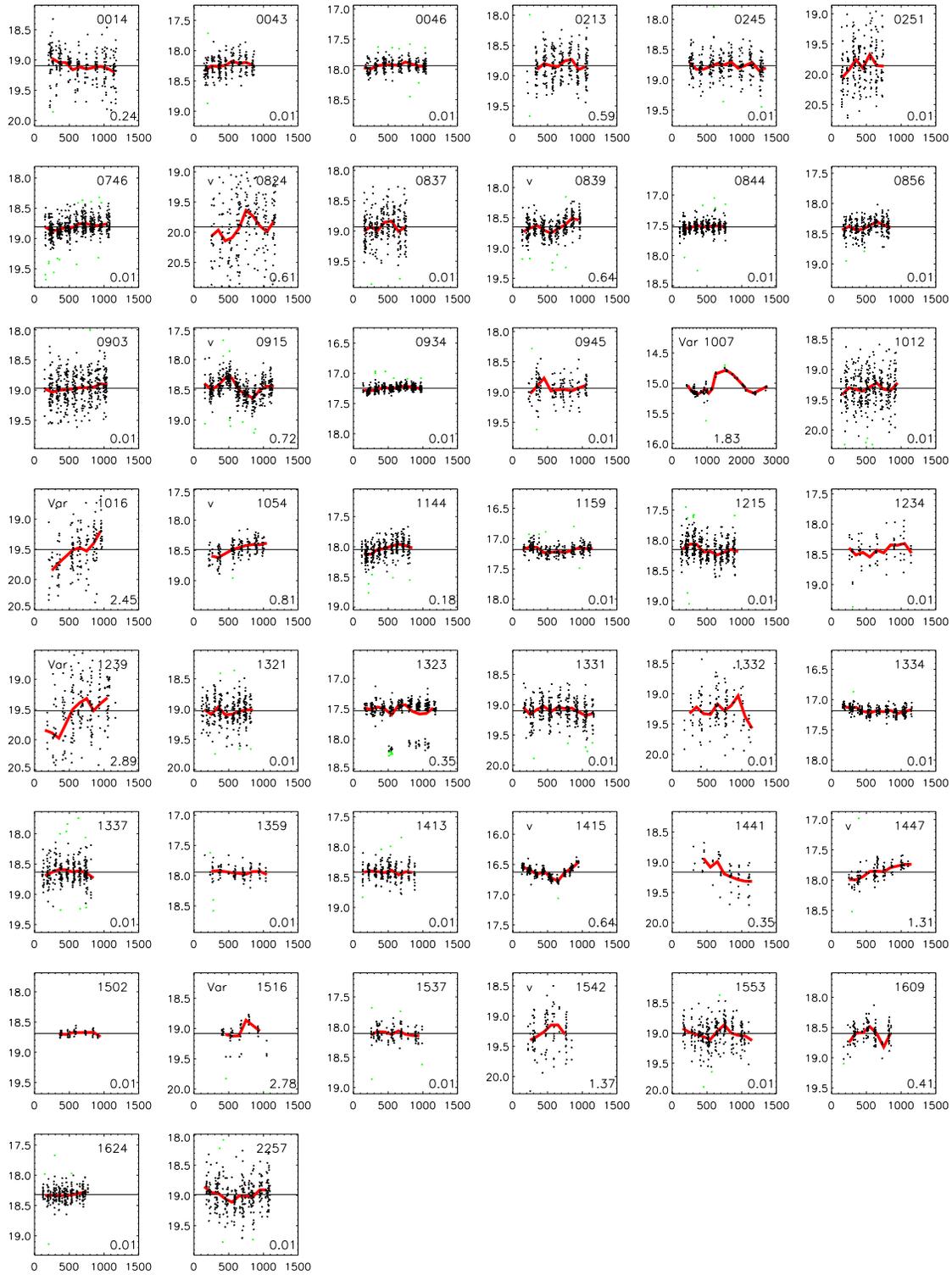} 
\caption{\small Optical continuum magnitude for BAL RLQs (labeled by
  truncated RA), plotted versus rest-frame timescale where the zero
  point is at MJD 53000. The red line is a running mean within
  rest-frame 200~d bins after outlier rejection (green points omitted,
  black points retained). A measure of variability is printed at lower
  right for each object; see $\S$4.2 for details. Objects with mild
  and strong variability are labeled as v and Var, respectively.}
\end{figure*}

BALs in RLQs tend to vary within only a fraction of the full velocity
width of the BAL trough, similar to RQQs (Gibson et al.~2008; G10;
C11). We define $v_{\rm BAL}$ to be the velocity span calculated from
the wavelength edges of the BAL, as defined in $\S$3.2 and listed in
Table~3, and $v_{\rm var}$ to be the velocity span of the variable
portion within the BAL. Figure~12 plots the velocity widths $v_{\rm
  var}$ and $v_{\rm var}/v_{\rm BAL}$ against $v_{\rm BAL}$, for those
RLQs with significant BAL variability and for RQQs from C11. Here the
open symbols show the segments of a given trough that varied (three
RQQs with variability within two distinct troughs are connected with
lines). For RLQs the longest separation measurement is plotted with
larger symbol size, while for RQQs only the longest separation
measurement is plotted for clarity. For both RLQs and RQQs the
velocity width of the varying regions tends to be only a few thousand
km~s$^{-1}$, as previously found for RQQs by Gibson et
al.~(2008). These and the previous results are consistent with a
simple scenario in which component segments within a given BAL have a
uniform and independent probability to vary, as could arise from
moving material at different radial velocities passing transversely
through the line of sight.

It would also be of interest to compare and contrast variability
within the \ion{Si}{IV} absorption region to that discussed here for
\ion{C}{IV} BALs. Unfortunately, our sample is selected in $z$ for
\ion{C}{IV} coverage and several of these BAL RLQs do not have
\ion{Si}{IV} coverage, which makes a robust statistical comparison
difficult. In RQQs, \ion{Si}{IV} absorption may be more variable than
\ion{C}{IV} (Capellupo et al.~2012; but see also G10) and segments at
similar velocities may show coordinated variability (G10; Capellupo et
al.~2012); a larger sample (restricted to higher redshifts) could test
whether this also holds for RLQs.

\subsection{Optical continuum variability}

We next investigate continuum variability in BAL quasars. One
motivation for considering optical continuum variability is that it
could be indicative of direct incident flux altering the absorber
ionization state (e.g., Trevese et al.~2013) or covering
factor. However, such potential connections are better explored with
EUV or X-ray observations (e.g., Gallagher et al.~2004; Saez et
al.~2012; Hamann et al.~2013), which directly probe the high-energy
radiation relevant to BAL shielding and driving. Of greater relevance,
a mutual origin of BAL outflows and optical emission in the accretion
disk might link continuum and absorption line variability; for
example, a particularly inhomogeneous disk in some quasars (e.g.,
speculatively due to near-Eddington accretion) might facilitate the
launching of absorbing clumps that are then observable as BALs.

Optical magnitudes at multiple epochs were obtained from the Catalina
Sky Survey{\footnote{\tt http://nessi.cacr.caltech.edu/DataRelease/}}
(Drake et al.~2009), Data Release 2. These unfiltered CCD measurements
of the quasar optical continua are plotted for BAL RLQs in Figure~13
and for BAL RQQs in Appendix~B. The magnitudes are screened for
outliers using two passes of 3$\sigma$ rejection, after which the
median magnitude is taken as the baseline brightness (dashed line in
each frame). Timescales are converted to rest-frame with MJD 53000 as
the fixed zero point (for reference, the DR2 release date for SDSS is
53079). The plots show a running mean calculated within rest-frame
timescale bins of 200~d for bins containing at least 4 valid
measurements (plotted as a solid red line).

We quantify optical continuum variability using a structure function,
following the general approach of Rengstorf et al.~(2006; see also
Vanden Berk et al.~2004; di Clemente et al.~1996). In addition to
comparing optical continuum variability in BAL RLQs versus BAL RQQs,
we wish to test whether optical continuum variability is linked to BAL
variability. Given the high-cadence but irregular monitoring of the
Catalina Sky Survey and the desired application of a uniform procedure
for assessing variability within each BAL quasar, we choose to bin the
$\sigma$-clipped magnitudes within intervals of 20 rest-frame days
(here we are only interested in variability on longer timescales)
prior to calculating the structure function. Errors within each bin
are estimated including both the provided measurement uncertainties
and the empirical scatter and then additionally enhanced by 30\%. This
slight initial smoothing and conservative inflation of errors does not
impact the relative ranking between individual objects or the BAL RLQs
versus BAL RQQs comparison but may give somewhat lower absolute
structure function values than other approaches. The structure
function is then considered for time lags up to 1000 rest-frame days,
binned by 100 days. For each individual quasar, the sum of these 10
measurements (or weighted for truncated coverage) is used to quantify
optical continuum variability, and these values are listed in
Figure~13 and Appendix~B.

\begin{figure}
\includegraphics[scale=0.45]{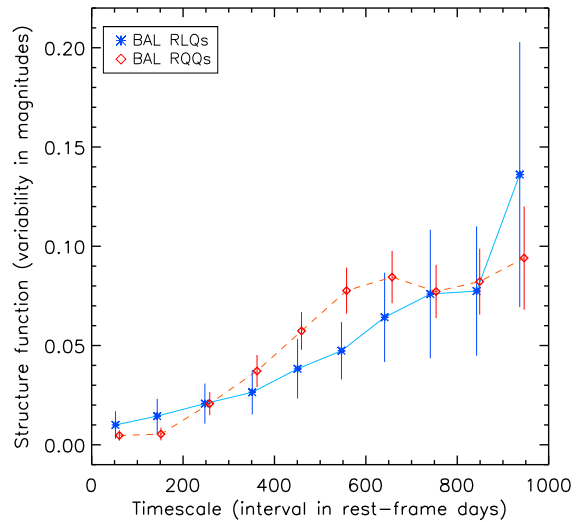} 
\caption{\small Optical continuum variability as a function of time
  lag for BAL RLQs (blue stars) and BAL RQQs (red diamonds). The
  structure function is calculated as detailed in $\S$4.2. Both BAL
  RLQs and RQQs show greater variability on longer timescales, and
  tend toward similar variability for any given interval.}
\end{figure}

\begin{figure}
\includegraphics[scale=0.45]{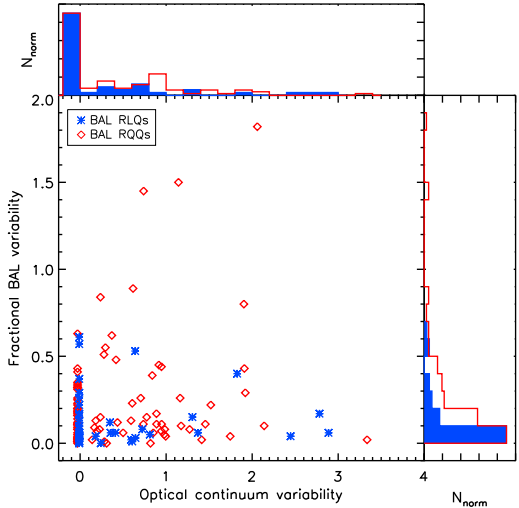} 
\caption{\small Optical continuum variability versus BAL variability
  for BAL RLQs (blue stars) and BAL RQQs (red diamonds). There is no
  apparent correlation. The one-dimensional variability distributions
  (shown as peak-normalized histograms) are similar, with BAL RLQs
  perhaps slightly less variable by both metrics.}
\end{figure}

The structure functions for BAL RLQs and for BAL RQQs averaged across
objects (rather than across raw magnitude measurements) are given in
Figure~14. The previously known tendency for quasars to display
greater variability on longer timescales (e.g., above references) is
clearly also present for BAL quasars, both RLQs and RQQs. There does
not appear to be a significant difference between BAL RLQs and BAL
RQQs in optical continuum variability (see Figure~14; if anything, BAL
RLQs may be less variable). The mean continuum variability for BAL
RLQs (for consistency again filtering out objects with
${\langle}EW{\rangle}<3.5$~\AA) is $0.37\pm0.11$, similar to the value
of $0.47\pm0.07$ for BAL RQQs. From 41 BAL RLQs, 7 (3) or 17\% (7\%)
show mild (strong) continuum variability; in comparison, from 85 BAL
RQQs, 22 (7) or 26\% (8\%) show mild (strong) continuum
variability. We find no significant correlation between optical
continuum and BAL variability (Table~6 and Figure~15).

Several previous studies have identified a tendency for RLQs to be
more variable than are RQQs (e.g., Vanden Berk et al.~2004; Garcia et
al.~1999; and references therein). We speculate that those samples
include a substantial number of RLQs for which the inclination is
close to the line of sight (including blazars) and that some of the
variability in such RLQs may be jet-linked. The similarity between BAL
RLQs and BAL RQQs might then support some geometric dependence to BAL
outflows (i.e., our RLQs are not viewed down the jet). This is
consistent with the intranight optical variability analysis of Joshi
\& Chand (2013), which found similar low variability in BAL RLQs and
BAL RQQs. Regardless of the underlying physical explanation, the
optical continuum results further support that variability in BAL RLQs
is similar to (or modestly less than) that in BAL RQQs.

\subsection{Influence of radio properties}

The absolute change and absolute fractional change in equivalent width
are plotted versus radio luminosity and radio loudness in
Figure~16. No strong dependencies of BAL variability upon radio
properties are apparent. The median and mean values of $|{\Delta}EW|$,
or $|{\Delta}EW|/{\langle}EW{\rangle}$ are similar for RLQs split at
either ${\ell}_{\rm r}=33$ or $R^{*}=2$, and KS tests find no
significant differences in their distributions (Table~5). This is
confirmed by Kendall and Spearman correlation tests (Table~6), which
show no significant correlation (probability $<0.5$ in all cases) of
$|{\Delta}EW|$ or $|{\Delta}EW|/{\langle}EW{\rangle}$ with either
${\ell}_{\rm r}$ or $R^{*}$. At most there is a very slight tendency,
not statistically significant, for increased
$|{\Delta}EW|/{\langle}EW{\rangle}$ toward higher $R^{*}$
values. However, this may be influenced by the relatively large
(within our sample) $|{\Delta}EW|/{\langle}EW{\rangle}$ values of a
few lobe-dominated quasars, for which the $R^{*}$ values tend to be
high (Figure~16) and might indeed be somewhat overestimated due to
inclusion of lobe emission or intrinsic reddening depressing the
optical continuum. Groupings of RLQs divided by ${\Delta}\tau$ or
${\langle}EW{\rangle}$ also do not show any significant trends, with
the exception of an apparent anti-correlation between $|{\Delta}EW|$
and $R^{*}$ for RLQs with ${\langle}EW{\rangle}>20$~\AA~that may again
be related to sample inhomogeneity (in this case, two varying RLQs
that have large ${\Delta}\tau$ timescales).

It may be noted from Figure~16 that the lobe-dominated RLQs tend
toward greater fractional variability than the core-dominated
RLQs. Indeed, the mean $|{\Delta}EW|/{\langle}EW{\rangle}$ is
$0.24\pm0.07$ ($0.09\pm0.02$) for lobe-dominated (core-dominated)
RLQs, and KS tests support marginal ($p=0.04$) inconsistency. However,
the small number of lobe-dominated BAL RLQs in our sample, as well as
their generally greater ${\Delta}\tau$ and smaller
${\langle}EW{\rangle}$ values (median 1500~d versus 600~d and
11~\AA~versus 20~\AA, respectively), indicates additional study is
required to confirm these conclusions. Anecdotally, other cases of
notable BAL variability in lobe-dominated RLQs are known; for example,
Hall et al.~(2011) report dramatic variability in the Mg~II and Fe~II
absorption features in the lobe-dominated RLQ FBQS J1408+3054 (the
redshift of $z=0.848$ precludes optical coverage of the \ion{C}{IV}
region; this is a ``FeLoBAL'' object that shows absorption within
lower ionization features, in this case including iron).

Studies of the radio spectral indices of non-BAL and BAL RLQs have
found that BAL RLQs tend to have steeper values of ${\alpha}_{\rm r}$,
suggestive of greater inclinations to the line of sight (DiPompeo et
al.~2012; Bruni et al.~2012). This is consistent with a geometrical
dependence to BAL structure, although it does appear that outflows can
exist at equatorial-to-polar latitudes. Within our sample, there is no
strong dependence between ${\alpha}_{\rm r}$ and BAL variability in
core-dominated BAL RLQs; lobe-dominated RLQs, with generally steep
radio spectral indices, may tend toward somewhat greater absorption
variability as discussed above.

\subsection{Relevance to outflow models}

In a disk-wind scenario, outflows launched from a rotating disk could
maintain an approximately Keplerian transverse velocity while
traveling radially, and consequent changes in the covering factor as
clouds move across the (extended) source can provide an explanation
for the observed minor shifts in depths at constant line-of-sight
velocity that characterize BAL variability in RQQs (Gibson et
al.~2008; G10; Capellupo et al.~2012). If lobe-dominated RLQs, known
to be more inclined than core-dominated RLQs, indeed show enhanced BAL
variability, then (particularly given the lack of correlation between
variability and general radio properties) this requires some
geometrical dependence of the BAL outflow structure. The very presence
of BALs in flat-spectrum, core-dominated RLQs is likely incompatible
with a strictly equatorial outflow\footnote{Note, however,
  determination of BAL RLQs as possessing polar outflows based solely
  on radio variability and inferred brightness temperature may be
  problematic (Hall \& Chajet~2011).}, but simulations of line-driven
disk winds indicate that material may be ejected at a range of angles
relative to the accretion disk (e.g., Giustini \& Proga 2012). For the
BAL RLQs considered here, it appears unnnecessary to invoke an
evolutionary phase, in which the quasar is nearly completely
enshrouded, to explain the presence of BALs. Recall, however, that our
sample is composed almost exclusively of HiBALs, and LoBALs may have
distinct properties ($\S$1; White et al.~2007).

\begin{figure}
\includegraphics[scale=0.38]{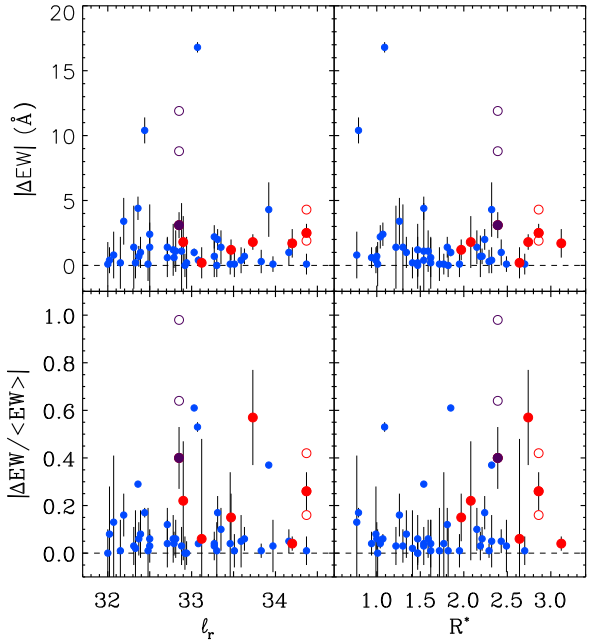}
\caption{\small Absolute change in BAL equivalent width and absolute
  fractional change in BAL equivalent width versus radio luminosity
  (left) and radio loudness (right) for BAL RLQs.  Core-dominated and
  lobe-dominated objects are plotted as smaller blue and larger red
  circles, respectively (with the lobe-dominated PG~1004+130 in
  purple; here and for SDSS J004323.43−001552.4 shorter-separation
  measurements are shown as open circles). There are no obvious strong
  trends with ${\ell}_{\rm r}$ or $R^{*}$. The lobe-dominated RLQs
  tend to display greater absolute and fractional variability than the
  core-dominated objects. }
\end{figure}

The lack of any apparent correlation between BAL variability and (core
plus lobe) radio loudness or luminosity would seem to suggest that the
strength of the jet does not exercise a controlling influence upon the
absorbing outflow. While the scarcity of BAL RLQs with both high
values of $R^{*}$ and large \ion{C}{IV} absorption $EW$ (Figure~1)
could indicate a physical connection (see also Shankar et al.~2008 for
modeling of the radio-loud and BAL fractions), it might alternatively
be due simply to a low likelihood for any given object to possess,
independently, extreme radio and BAL properties. If the jet and wind
are not intimately connected in BAL RLQs, this provides an interesting
contrast with the situation for \hbox{X-ray} binaries, for which it is
found that the jet-aided development of a radiation-driven wind can
remove sufficient material to starve a jet (e.g., Neilsen \& Lee
2009), in a feedback cycle between the ``low/hard'' and ``high/soft''
states. The longer timescales (scaling with black hole mass) in
quasars, perhaps in concert with a greater influence of the corona
upon the accretion structure than operates in \hbox{X-ray} binaries,
appear to permit dual-mode feedback with both the high-velocity,
low-mass jets and the relatively lower-velocity, higher-mass winds
(e.g., Proga et al.~2010) capable of significant energy injection into
their surroundings. If indeed mechanical power is ejected primarily in
the form of jets below $L_{\rm bol}\sim10^{-2}L_{\rm Edd}$ and as
winds at higher accretion luminosities (King et al.~2013), RLQs may
sit near this boundary. RLQs also hosting BALs\footnote{If BALs in
RLQs are only detectable along a particular line of sight, the
intrinsic fraction of RLQs hosting BALs could be much larger than
observed.} might therefore be expected to be particularly efficient at
quenching star-formation within their host galaxies.

\section{Summary}

We have carried out a study of \ion{C}{IV} BAL variability within a
representative sample of 46 RLQs, using 78 pairs of spectra collected
with rest-frame separations of 80--6000~d. Our primary results are the
following:

1. Changes in BAL structure are generally either modest or
statistically insignificant; the mean absolute fractional change
$|{\Delta}EW|/{\langle}EW{\rangle}$ is $0.10\pm0.02$, somewhat lower
than the $0.17\pm0.03$ found in a matched sample of BAL RQQs. BAL
variability in RLQs, where observed, typically consists of a slight
change in the depth of a portion of the absorption trough,
qualitatively similar to BAL RQQ variability (see $\S$3.2, 4.1,
Table~5, Appendix~A).

2. BALs in RLQs vary more on longer timescales (as do BALs in
RQQs). The absolute fractional change in equivalent width is
correlated with ${\Delta}\tau$, and the mean
$|{\Delta}EW|/{\langle}EW{\rangle}$ is $0.13\pm0.03$ ($0.06\pm0.02$)
for ${\Delta}\tau\ge500$~d ($<500$~d) (see $\S$4.1, Table~5,
Figure~10).

3. BALs of larger equivalent width tend to show greater absolute (but
not fractional) changes in equivalent width, for both RLQs and RQQs,
perhaps due to their coverage of more discrete velocity segments (see
$\S$4.1, Table~6, Figure~11).

4. BAL RLQs and BAL RQQs show similar optical continuum variability
patterns (if anything, BAL RLQs are less variable) and both vary more
on longer timescales, in common with non-BAL quasars (see $\S$4.2,
Figures~13--14, Appendix B).

5. Radio loudness and luminosity do not influence BAL variability. The
mean $|{\Delta}EW|$ and $|{\Delta}EW|/{\langle}EW{\rangle}$ values for
RLQs divided at $R^{*}=2$ or ${\ell}_{\rm r}=33$ are similar, and
neither measure of BAL variability is significantly correlated with
either $R^{*}$ or ${\ell}_{\rm r}$ for these BAL RLQs (see $\S$4.3,
Table~6).

6. Lobe-dominated RLQs apparently show greater fractional BAL
variability than do core-dominated RLQs: the mean
$|{\Delta}EW|/{\langle}EW{\rangle}$ is $0.24\pm0.07$ ($0.09\pm0.02$)
for lobe-dominated (core-dominated) RLQs. However, the small number of
lobe-dominated BAL RLQs in our sample, as well as their generally
greater ${\Delta}\tau$ and smaller ${\langle}EW{\rangle}$ values here,
indicates additional study is warranted to confirm these conclusions
(see $\S$4.3, Figure~16). 

In summary, BAL variability in RLQs appears to occur in a similar
manner as in RQQs (consistent with a common physical mechanism, such
as a disk wind), but at perhaps weaker amplitude or with lesser
frequency. The lack of dependence upon radio properties suggests that
the jet in RLQs does not strongly influence the absorber, while the
potential greater variability in lobe-dominated objects supports some
geometrical dependence to the outflow structure.

\section{Acknowledgments}

We thank Mike Eracleous, Karen Lewis, and Paola Rodriguez Hidalgo for
helpful conversations and suggestions, Mike Brotherton for providing a
spectrum for J1016, and Paul Martini for assistance with OSMOS. We
thank the anonymous referee for a thoughtful and constructive report
that improved this paper. This work was partially conducted by CAW
through the Research Experiences for Undergraduates program at the
College of Wooster. WNB acknowledges NSF grant AST 1108604 and NASA
ADP grant NNX10AC99G.

This project made use of the NRAO VLA Archive Survey, (c)
AUI/NRAO. The NVAS can currently be browsed through
http://www.aoc.nrao.edu/$\sim$vlbacald/.

This research has made use of the NASA/IPAC Extragalactic Database
(NED) which is operated by the Jet Propulsion Laboratory, California
Institute of Technology, under contract with the National Aeronautics
and Space Administration.

The CSS survey is funded by the National Aeronautics and Space
Administration under Grant No. NNG05GF22G issued through the Science
Mission Directorate Near-Earth Objects Observations Program.  The CRTS
survey is supported by the U.S.~National Science Foundation under
grants AST-0909182.

Funding for the SDSS and SDSS-II has been provided by the Alfred
P. Sloan Foundation, the Participating Institutions, the National
Science Foundation, the U.S. Department of Energy, the National
Aeronautics and Space Administration, the Japanese Monbukagakusho, the
Max Planck Society, and the Higher Education Funding Council for
England. The SDSS Web Site is http://www.sdss.org/.

The Hobby-Eberly Telescope (HET) is a joint project of the University
of Texas at Austin, the Pennsylvania State University, Stanford
University, Ludwig-Maximillians-Universit\"at M\"unchen, and
Georg-August-Universit\"at G\"ottingen. The HET is named in honor of
its principal benefactors, William P. Hobby and Robert E. Eberly. The
Marcario Low-Resolution Spectrograph is named for Mike Marcario of
High Lonesome Optics, who fabricated several optics for the instrument
but died before its completion; it is a joint project of the
Hobby-Eberly Telescope partnership and the Instituto de
Astronom\'{\i}a de la Universidad Nacional Aut\'onoma de M\'exico.

\begin{table*}
\centering
\scriptsize
\begin{minipage}{17.5cm}
\caption{BAL RLQs: radio and optical properties and longest-separation summed BAL measurements}
\begin{tabular}{p{70pt}rrrrrrrrrrrl}
\hline 
Name (SDSS J2000) & $z$ & $m_{\rm i}$ & $M_{\rm i}$ & ${\ell}_{\rm
  uv}$\footnote{Monochromatic optical/UV or radio luminosities are
  given at rest-frame 2500~\AA~or 5~GHz, respectively, in units of
  erg~s$^{-1}$~Hz$^{-1}$, expressed as a logarithm. The radio loudness
  is calculated as $R^{*}={\ell}_{\rm r}-{\ell}_{\rm uv}$.} &
  ${\ell}_{\rm r}$ & $R^{*}$ & ${\Delta}(g-i)$\footnote{The relative
  color ${\Delta}(g-i)$ is calculated as $g-i$ less the median value
  for quasars near that redshift. Redder objects have
  ${\Delta}(g-i)>0$.} & ${\Delta}\tau$\footnote{Rest-frame time
  between BAL measurements, in days.} & ${\langle}EW{\rangle}$\footnote{Average equivalent width, in~\AA.} &
  ${\Delta}EW$\footnote{Change in equivalent width, in~\AA.} &
  $\frac{|{\Delta}EW|}{{\langle}EW{\rangle}}$\footnote{Absolute
  fractional change in equivalent width} & Notes\footnote{Lo: known
  LoBAL; MB: mini-BAL present; LD: lobe-dominated; J: small-scale
  milliarcsecond jet; FS: flat radio spectrum} \\ 
\hline
001438.28$-$010750.1 & 1.806 &  18.36 &  $-$26.77 &30.99 &  32.00 &   1.01 &    0.59  &   258 &   16.2 &      $-$0.1$\pm$1.7  &   0.00	& \\                     
004323.43$-$001552.4 & 2.798 &  18.15 &  $-$27.96 &31.51 &  34.37 &   2.86 &    0.38  &   675 &    9.4 &      $-$2.5$\pm$0.7  &   0.26	& LD \\ 		      
004613.54$+$010425.7 & 2.152 &  17.77 &  $-$27.76 &31.43 &  32.50 &   1.07 &    0.31  &  1881 &   39.5 &      $-$2.4$\pm$0.9  &   0.06	& \\ 		      
021333.34$+$003030.8 & 2.053 &  18.90 &  $-$26.52 &30.92 &  32.33 &   1.41 &    0.80  &   854 &    7.6 &      $-$0.2$\pm$1.2  &   0.02	& \\ 		      
024534.07$+$010813.7 & 1.536 &  18.37 &  $-$26.38 &30.76 &  32.48 &   1.72 &    0.67  &   436 &    7.5 &      $+$0.1$\pm$1.3  &   0.01	& \\ 		      
025105.14$-$001732.1 & 3.456 &  19.53 &  $-$27.05 &31.30 &  32.07 &   0.77 &    0.66  &    69 &    6.4 &      $+$0.8$\pm$1.8  &   0.13	& \\ 		      
074610.50$+$230710.7 & 2.093 &  18.27 &  $-$27.20 &31.20 &  33.35 &   2.15 &    1.06  &   624 &   14.5 &      $+$1.4$\pm$1.3  &   0.10	& FS \\ 		      
082450.19$+$440212.8 & 1.866 &  19.59 &  $-$25.61 &30.53 &  32.15 &   1.62 &    0.25  &    87 &   15.0 &      $+$0.2$\pm$2.0  &   0.01	& \\ 		      
083749.59$+$364145.5 & 3.416 &  18.54 &  $-$28.02 &31.54 &  33.83 &   2.29 &    0.61  &   496 &   33.2 &      $-$0.3$\pm$0.9  &   0.01	& \\ 		      
083925.61$+$045420.2 & 2.447 &  18.30 &  $-$27.52 &31.34 &  32.88 &   1.54 &    1.80  &    17 &   44.8 &      $-$1.1$\pm$3.7  &   0.03	& Lo \\ 		      
084401.95$+$050357.9 & 3.346 &  17.01 &  $-$29.50 &32.23 &  33.27 &   1.04 &    1.10  &   420 &   57.1 &      $+$2.2$\pm$0.9  &   0.04	& \\ 		      
085641.56$+$424253.9 & 3.062 &  18.42 &  $-$27.90 &31.42 &  33.63 &   2.21 &    0.07  &   546 &   11.2 &      $-$0.7$\pm$0.6  &   0.06	& FS \\ 		      
090306.67$+$272532.9 & 2.233 &  18.86 &  $-$26.75 &31.05 &  32.39 &   1.34 &    0.45  &   319 &   11.7 &      $+$1.0$\pm$1.2  &   0.08	& \\ 		      
091512.53$+$305014.9 & 1.983 &  18.54 &  $-$26.80 &31.03 &  32.02 &   0.99 &    0.35  &   130 &    4.8 &      $-$0.4$\pm$1.0  &   0.08	& \\ 		      
093403.96$+$315331.3 & 2.422 &  17.07 &  $-$28.73 &31.85 &  32.79 &   0.94 &    0.80  &   300 &   17.2 &      $+$0.6$\pm$0.8  &   0.04	& \\ 		      
094513.89$+$505521.8 & 2.137 &  18.59 &  $-$26.93 &31.09 &  32.31 &   1.22 &    0.97  &   672 &   42.8 &      $-$1.4$\pm$3.2  &   0.03	& Lo \\ 		      
100726.10$+$124856.2 & 0.241 &  15.17 &  $-$25.12 &30.46 &  32.85 &   2.39 &  $-$0.51 &  6163 &    7.8 &      $-$3.1$\pm$1.0  &   0.40	& LD \\ 		      
101219.48$+$350333.8 & 2.628 &  19.02 &  $-$26.96 &31.20 &  32.50 &   1.30 &    0.74  &   114 &   50.1 &      $-$1.4$\pm$3.3  &   0.03	& \\ 		      
101614.26$+$520915.7 & 2.460 &  19.23 &  $-$26.62 &31.08 &  34.20 &   3.12 &    1.39  &  1198 &   41.1 &      $-$1.7$\pm$1.1  &   0.04	& LD \\		      
105416.51$+$512326.0 & 2.340 &  18.47 &  $-$27.25 &31.27 &  33.59 &   2.32 &    0.35  &   549 &    9.1 &      $+$0.4$\pm$1.0  &   0.05	& FS \\ 		      
114436.65$+$095904.9 & 3.150 &  18.07 &  $-$28.31 &31.69 &  33.46 &   1.77 &    0.21  &   419 &    2.5 &      $-$0.1$\pm$0.8  &   0.04	& \\ 		      
115944.82$+$011206.9 & 2.002 &  16.98 &  $-$28.39 &31.67 &  34.37 &   2.70 &    0.40  &   952 &   12.7 &      $+$0.1$\pm$0.8  &   0.01	& J, MB? \\ 	      
121539.66$+$090607.4 &  2.723 & 18.09 &  $-$27.97 &31.60 &  33.92 &   2.32 &    0.18  &	1571 &   11.6 &      $-$4.3$\pm$2.1  &   0.37	& \\		      
123411.74$+$615832.5 & 1.949 &  18.37 &  $-$26.93 &31.08 &  33.27 &   2.19 &    0.46  &   745 &   28.6 &      $-$0.7$\pm$1.6  &   0.03	& \\ 		      
123954.15$+$373954.5 & 1.841 &  19.70 &  $-$25.47 &30.48 &  33.12 &   2.64 &    0.41  &   770 &    2.8 &      $+$0.2$\pm$1.2  &   0.06	& LD, Lo \\ 	      
132139.86$-$004151.9 & 3.074 &  18.66 &  $-$27.66 &31.47 &  32.94 &   1.47 &    1.46  &    77 &   39.9 &      $-$0.2$\pm$2.0  &   0.00	& Lo \\ 		      
132304.58$-$003856.5 & 1.827 &  17.81 &  $-$27.34 &31.22 &  32.81 &   1.59 &    0.53  &   113 &   19.3 &      $-$1.1$\pm$1.6  &   0.06	& \\ 		      
133150.51$+$004518.8 &  1.885 & 18.89 &  $-$26.33 &30.82 &  32.36 &   1.54 &    0.15  &	 101 &   15.4 &      $+$4.4$\pm$0.9  &   0.29	& \\		      
133259.16$+$490946.9 & 1.995 &  19.08 &  $-$26.27 &30.82 &  32.90 &   2.08 &    1.15  &   503 &    8.0 &      $+$1.8$\pm$2.0  &   0.22	& LD \\ 		      
133428.06$-$012349.0 &  1.876 & 16.80 &  $-$28.41 &31.65 &  32.44 &   0.79 &    1.02  &	2027 &   60.3 &      $-$10.4$\pm$1.0 &   0.17	& Lo\\		      
133701.39$-$024630.2 & 3.063 &  18.41 &  $-$27.91 &31.48 &  33.97 &   2.49 &    0.18  &   512 &    5.1 &      $-$0.1$\pm$0.6  &   0.03	& FS \\ 		      
135910.45$+$563617.3 & 2.248 &  17.76 &  $-$27.87 &31.49 &  33.30 &   1.82 &    0.58  &   933 &    7.6 &      $+$0.0$\pm$0.9  &   0.01	& \\ 		      
141151.97$+$550948.7 & 1.889 &  18.47 &  $-$26.76 &30.99 &  33.73 &   2.74 &    0.18  &   929 &    3.1 &      $-$1.8$\pm$0.6  &   0.57	& LD \\ 		      
141334.38$+$421201.7 & 2.818 &  18.23 &  $-$27.91 &31.56 &  33.51 &   1.95 &    0.53  &   465 &   14.5 &      $+$0.1$\pm$1.0  &   0.01	& J, FS \\ 	      
141546.24$+$112943.4 & 2.560 &  16.91 &  $-$29.02 &31.98 &  33.07 &   1.09 &    0.29  &  1954 &   31.9 &      $-$16.8$\pm$0.4 &   0.53	& Lo \\ 		      
144136.54$+$632519.4 & 1.779 &  18.55 &  $-$26.54 &30.90 &  32.71 &   1.81 &    0.10  &   792 &   11.7 &      $-$1.4$\pm$0.8  &   0.12	& \\ 		      
144707.40$+$520340.1 & 2.065 &  17.47 &  $-$27.96 &31.50 &  33.47 &   1.97 &    0.44  &   486 &    7.6 &      $+$1.2$\pm$0.8  &   0.15	& LD \\ 		      
150206.66$-$003606.9 &  2.200 & 18.47 &  $-$27.11 &31.18 &  33.03 &   1.85 &    0.15  &   118 &    1.7 &      $+$1.0$\pm$0.3  &   0.61	& \\		      
151630.30$-$005625.5 & 1.921 &  18.31 &  $-$26.96 &31.08 &  33.31 &   2.24 &    0.39  &  1131 &   11.7 &      $-$2.0$\pm$0.8  &   0.17	& \\ 		      
153703.95$+$533220.0 &  2.406 & 17.90 &  $-$27.88 &31.54 &  33.08 &   1.54 &    0.20  &    20 &   10.3 &      $+$0.4$\pm$0.3  &   0.04	& \\		      
154241.14$+$532204.3 & 2.983 &  19.11 &  $-$27.15 &31.31 &  32.78 &   1.47 &    1.08  &    17 &   20.3 &      $-$1.2$\pm$2.0  &   0.06	& \\ 		      
155355.38$+$324513.3 & 2.065 &  18.90 &  $-$26.54 &30.93 &  32.19 &   1.26 &    0.44  &   473 &   21.0 &      $-$3.4$\pm$1.8  &   0.16	& \\ 		      
160943.33$+$522550.9 & 2.723 &  18.59 &  $-$27.47 &31.37 &  32.37 &   1.00 &    0.38  &   588 &   12.1 &      $+$0.7$\pm$0.8  &   0.06	& \\ 		      
162453.47$+$375806.6 & 3.380 &  18.17 &  $-$28.37 &31.73 &  34.16 &   2.43 &    0.22  &   396 &   20.6 &      $-$1.0$\pm$1.0  &   0.05	& J, MB \\ 	      
225706.17$-$002532.8 & 1.985 &  18.40 &  $-$26.95 &31.09 &  32.71 &   1.62 &    1.11  &  1318 &   14.3 &      $-$0.6$\pm$1.5  &   0.04	& \\ 		      
235702.54$-$004824.0 & 3.013 &  18.68 &  $-$27.61 &31.45 &  32.92 &   1.47 &    0.42  &   103 &   12.8 &      $-$0.0$\pm$0.9  &   0.00   & \\ 		      
\hline
\end{tabular}
\end{minipage}
\end{table*}
\clearpage

\begin{table*}
\centering
\scriptsize
\begin{minipage}{13.0cm}
\caption{Measurements of RLQ \ion{C}{IV} absorption troughs}
\begin{tabular}{p{20pt}rrrlcp{20pt}rrrl}
\hline
Object\footnote{Truncated, see Table~1 for full name.} &
${\lambda}_{1}-{\lambda}_{2}$\footnote{Wavelength boundaries (in~\AA) of BAL;
  see $\S$3.2 for details.}  & MJD & $EW$~(\AA) &
Obs\footnote{Observations from S: Sloan Digital Sky Survey; H: Hobby
  Eberly Telescope; I: International Ultraviolet Explorer; T: Hubble
  Space Telescope; M: MDM Observatory} & & Object &
${\lambda}_{1}-{\lambda}_{2}$ & MJD & $EW$~(\AA) & Obs \\
\hline
0014 &  1525--1548 & 51795 &  16.29$\pm$1.18 &  S  &~&  1012 &  1440--1545 & 53357 &  50.76$\pm$2.98 &  S \\   
     &             & 52518 &  16.21$\pm$1.18 &  S  &~&       &             & 53771 &  49.35$\pm$1.36 &  S \\   
0043 &  1444--1499 & 51794 &  10.61$\pm$0.61 &  S  &~&  1016 &  1509--1554 & 51546 &  28.69$\pm$0.54 &  K \\   
     &             & 52199 &  12.46$\pm$0.58 &  S  &~&       &             & 55684 &  29.35$\pm$0.71 &  H \\   
     &             & 54356 &   8.13$\pm$0.36 &  H  &~&       &  1460--1503 & 51546 &  13.32$\pm$0.24 &  K \\   
0046 &  1449--1493 & 51794 &  24.22$\pm$0.54 &  S  &~&       &             & 55684 &  10.93$\pm$0.55 &  H \\   
     &             & 52199 &  22.69$\pm$0.45 &  S  &~&  1054 &  1521--1543 & 52669 &   8.93$\pm$0.79 &  S \\   
     &             & 55776 &  22.29$\pm$0.32 &  H  &~&       &             & 54504 &   9.35$\pm$0.65 &  H \\   
     &  1524--1542 & 51794 &  15.54$\pm$0.80 &  S  &~&  1144 &  1473--1486 & 52734 &   2.60$\pm$0.51 &  S \\   
     &             & 52199 &  15.26$\pm$0.75 &  S  &~&       &             & 54472 &   2.49$\pm$0.58 &  H \\   
     &             & 55776 &  15.29$\pm$0.65 &  H  &~&  1159 &  1533--1549 & 51663 &  12.62$\pm$0.58 &  S \\  
0213 &  1535--1547 & 51816 &   7.72$\pm$0.99 &  S  &~&       &             & 51930 &  12.57$\pm$0.58 &  S \\  
     &             & 54413 &   7.55$\pm$0.68 &  H  &~&       &             & 54509 &  12.72$\pm$0.49 &  H \\  
0245 &  1536--1547 & 51871 &   7.40$\pm$0.91 &  S  &~&  1234 &  1463--1505 & 52373 &  28.98$\pm$1.30 &  S \\  
     &             & 52177 &   7.92$\pm$1.10 &  S  &~&       &             & 54552 &  28.61$\pm$0.72 &  H \\  
     &             & 52973 &   7.51$\pm$0.99 &  S  &~&       &             & 54579 &  28.24$\pm$0.96 &  H \\  
0251 &  1522--1534 & 51871 &   6.00$\pm$1.24 &  S  &~&  1321 &  1488--1549 & 51665 &  40.03$\pm$1.48 &  S \\  
     &             & 52177 &   6.82$\pm$1.30 &  S  &~&       &             & 51984 &  39.87$\pm$1.34 &  S \\  
0746 &  1525--1537 & 52577 &   5.65$\pm$0.72 &  S  &~&  1323 &  1495--1511 & 51665 &   8.76$\pm$0.78 &  S \\  
     &             & 54452 &   6.54$\pm$0.50 &  H  &~&       &             & 51984 &   8.39$\pm$0.58 &  S \\  
     &             & 54507 &   6.63$\pm$0.66 &  S  &~&       &  1518--1535 & 51665 &  11.08$\pm$0.93 &  S \\  
     &  1539--1549 & 52577 &   8.14$\pm$0.65 &  S  &~&       &             & 51984 &  10.35$\pm$0.85 &  S \\  
     &             & 54452 &   8.48$\pm$0.66 &  H  &~&  1332 &  1511--1530 & 52762 &   7.12$\pm$1.90 &  S \\  
     &             & 54507 &   8.54$\pm$0.65 &  S  &~&       &             & 54270 &   8.92$\pm$0.65 &  H \\  
0824 &  1518--1539 & 51959 &  14.89$\pm$1.49 &  S  &~&  1337 &  1528--1539 & 52427 &   5.13$\pm$0.42 &  S \\  
     &             & 52207 &  15.07$\pm$1.35 &  S  &~&       &             & 54504 &   5.00$\pm$0.36 &  H \\  
0837 &  1502--1543 & 52320 &  33.38$\pm$0.71 &  S  &~&  1359 &  1533--1549 & 52669 &   7.57$\pm$0.73 &  S \\  
     &             & 54413 &  32.91$\pm$0.83 &  H  &~&       &             & 55700 &   7.62$\pm$0.55 &  H \\  
     &             & 54452 &  33.57$\pm$0.69 &  H  &~&  1411 &  1534--1544 & 53088 &   3.96$\pm$0.55 &  S \\  
     &             & 54504 &  33.10$\pm$0.49 &  H  &~&       &             & 55775 &   2.20$\pm$0.24 &  H \\  
0839 &  1417--1461 & 52650 &  21.20$\pm$1.50 &  S  &~&  1413 &  1529--1549 & 52823 &  14.45$\pm$0.84 &  S \\  
     &             & 52708 &  21.05$\pm$2.39 &  S  &~&       &             & 54594 &  14.54$\pm$0.58 &  H \\  
     &  1497--1509 & 52650 &   7.44$\pm$0.83 &  S  &~&  1415 &  1476--1535 & 53848 &  34.65$\pm$0.68 &  S \\  
     &             & 52708 &   7.29$\pm$1.15 &  S  &~&       &             & 54567 &  26.73$\pm$0.70 &  M \\  
0839 &  1531--1554 & 52650 &  16.75$\pm$1.06 &  S  &~&       &             & 55688 &  30.61$\pm$0.47 &  H \\
     &             & 52708 &  15.91$\pm$1.56 &  S  &~&  1441 &  1526--1542 & 52339 &  12.40$\pm$0.67 &  S \\    
0844 &  1412--1529 & 52650 &  56.04$\pm$0.65 &  S  &~&       &             & 54533 &  10.95$\pm$0.50 &  S \\    
     &             & 54478 &  58.20$\pm$0.60 &  H  &~&  1447 &  1447--1487 & 52786 &   7.00$\pm$0.58 &  S \\    
0856 &  1508--1527 & 52294 &  11.61$\pm$0.48 &  S  &~&       &             & 54271 &   8.17$\pm$0.51 &  H \\    
     &             & 54414 &  10.86$\pm$0.43 &  H  &~&  1516 &  1532--1552 & 52404 &  12.69$\pm$0.64 &  S \\    
     &             & 54520 &  10.88$\pm$0.38 &  H  &~&       &             & 55707 &  10.73$\pm$0.51 &  H \\    
0903 &  1491--1515 & 53387 &  11.23$\pm$0.96 &  S  &~&  1542 &  1525--1554 & 52374 &  20.85$\pm$1.49 &  S \\    
     &             & 54416 &  12.21$\pm$0.71 &  H  &~&       &             & 52442 &  19.66$\pm$1.40 &  S \\    
0915 &  1535--1551 & 53379 &   5.01$\pm$0.75 &  S  &~&  1553 &  1476--1527 & 52825 &  22.67$\pm$1.56 &  S \\    
     &             & 53768 &   4.64$\pm$0.62 &  S  &~&       &             & 53227 &  19.42$\pm$1.26 &  S \\    
0934 &  1482--1499 & 53386 &   6.73$\pm$0.38 &  S  &~&       &             & 54270 &  19.25$\pm$0.86 &  H \\    
     &             & 54414 &   6.92$\pm$0.29 &  H  &~&  1609 &  1531--1547 & 52051 &  11.76$\pm$0.65 &  S \\    
     &  1406--1440 & 53386 &  10.15$\pm$0.48 &  S  &~&       &             & 54233 &  12.48$\pm$0.50 &  H \\    
     &             & 54414 &  10.60$\pm$0.48 &  H  &~&  1624 &  1403--1446 & 52767 &  15.32$\pm$0.55 &  S \\    
0945 &  1405--1500 & 52409 &  43.47$\pm$1.95 &  S  &~&       &             & 54270 &  14.20$\pm$0.42 &  H \\    
     &             & 54496 &  40.18$\pm$2.24 &  H  &~&       &             & 54504 &  14.25$\pm$0.49 &  H \\    
     &             & 54522 &  42.07$\pm$2.58 &  H  &~&       &  1528--1543 & 52767 &   5.82$\pm$0.47 &  S \\    
1007 &  1491--1532 & 45083 &   5.66$\pm$0.65 &  I  &~&       &             & 54270 &   5.17$\pm$0.38 &  H \\    
     &             & 46442 &  14.04$\pm$0.82 &  I  &~&       &             & 54504 &   5.85$\pm$0.40 &  H \\    
     &             & 52729 &   3.66$\pm$0.36 &  T  &~&  2257 &  1537--1554 & 51792 &  14.63$\pm$1.21 &  S \\    
     &  1537--1552 & 45083 &   3.67$\pm$0.52 &  I  &~&       &             & 52178 &  14.19$\pm$1.18 &  S \\    
     &             & 46442 &   4.09$\pm$0.46 &  I  &~&       &             & 55740 &  14.01$\pm$0.80 &  H \\    
     &             & 52729 &   2.54$\pm$0.31 &  T  &~&  2357 &  1525--1541 & 51791 &  12.82$\pm$0.58 &  S \\    
     &             &       &                 &     &~&       &             & 52203 &  12.77$\pm$0.70 &  S \\
\hline                                              
\end{tabular}                                       
\end{minipage}                                      
\end{table*}                                        
\clearpage

\begin{table*}
\centering
\scriptsize
\begin{minipage}{17.0cm}
\caption{Variability of RLQ \ion{C}{IV} absorption troughs}
\begin{tabular}{p{15pt}rrrrrcp{15pt}rrrrr}
\hline
Name\footnote{Truncated, see Table~1 for full name.} & ${\lambda}_{1}-{\lambda}_{2}$ & ${\Delta}\tau$  & ${\langle}EW{\rangle}$ & 
     ${\Delta}EW$ & $\frac{|{\Delta}EW|}{{\langle}EW{\rangle}}$ & &  
Name & ${\lambda}_{1}-{\lambda}_{2}$ & ${\Delta}\tau$ & ${\langle}EW{\rangle}$ & ${\Delta}EW$ & $\frac{|{\Delta}EW|}{{\langle}EW{\rangle}}$ \\
   & (\AA) & (d) & (\AA) & (\AA) & & & & (\AA) & (d) & (\AA) & (\AA) & \\
\hline
0014 & 1525--1548 &   258& 16.2&  $-$0.1$\pm$1.7 & 0.00$\pm$0.10 &~& 1007 & 1537--1552 &  6163&  3.1&  $-$1.1$\pm$0.6 & 0.36$\pm$0.20 \\
0043 & 1444--1499 &   675&  9.4&  $-$2.5$\pm$0.7 & 0.26$\pm$0.08 &~&      &            &  1095&  3.9&  $+$0.4$\pm$0.7 & 0.11$\pm$0.18 \\
     &            &   107& 11.5&  $+$1.9$\pm$0.8 & 0.16$\pm$0.07 &~&      &            &  5068&  3.3&  $-$1.6$\pm$0.6 & 0.47$\pm$0.17 \\
     &            &   568& 10.3&  $-$4.3$\pm$0.7 & 0.42$\pm$0.07 &~& 1012 & 1440--1545 &   114& 50.1&  $-$1.4$\pm$3.3 & 0.03$\pm$0.07 \\
0046 & 1449--1493 &  1264& 23.3&  $-$1.9$\pm$0.6 & 0.08$\pm$0.03 &~& 1016 & 1509--1554 &  1198& 29.0&  $+$0.7$\pm$0.9 & 0.02$\pm$0.03 \\
     &            &   129& 23.5&  $-$1.5$\pm$0.7 & 0.07$\pm$0.03 &~&      & 1460--1503 &  1198& 12.1&  $-$2.4$\pm$0.6 & 0.20$\pm$0.05 \\
     &            &  1136& 22.5&  $-$0.4$\pm$0.6 & 0.02$\pm$0.02 &~& 1054 & 1521--1543 &   549&  9.1&  $+$0.4$\pm$1.0 & 0.05$\pm$0.11 \\
     & 1524--1542 &  1264& 15.4&  $-$0.2$\pm$1.0 & 0.02$\pm$0.07 &~& 1144 & 1473--1486 &   419&  2.5&  $-$0.1$\pm$0.8 & 0.04$\pm$0.30 \\
     &            &   129& 15.4&  $-$0.3$\pm$1.1 & 0.02$\pm$0.07 &~& 1159 & 1533--1549 &   952& 12.7&  $+$0.1$\pm$0.8 & 0.01$\pm$0.06 \\
     &            &  1136& 15.3&  $+$0.0$\pm$1.0 & 0.00$\pm$0.06 &~&      &            &    89& 12.6&  $-$0.1$\pm$0.8 & 0.00$\pm$0.07 \\
0213 & 1535--1547 &   854&  7.6&  $-$0.2$\pm$1.2 & 0.02$\pm$0.16 &~&      &            &   862& 12.6&  $+$0.2$\pm$0.8 & 0.01$\pm$0.06 \\
0245 & 1536--1547 &   436&  7.5&  $+$0.1$\pm$1.3 & 0.01$\pm$0.18 &~& 1234 & 1463--1505 &   745& 28.6&  $-$0.7$\pm$1.6 & 0.03$\pm$0.06 \\
     &            &   121&  7.7&  $+$0.5$\pm$1.4 & 0.07$\pm$0.19 &~&      &            &   736& 28.8&  $-$0.4$\pm$1.5 & 0.01$\pm$0.05 \\
     &            &   315&  7.7&  $-$0.4$\pm$1.5 & 0.05$\pm$0.19 &~&      &            &     9& 28.4&  $-$0.4$\pm$1.2 & 0.01$\pm$0.04 \\
0251 & 1522--1534 &    69&  6.4&  $+$0.8$\pm$1.8 & 0.13$\pm$0.28 &~& 1321 & 1488--1549 &    77& 39.9&  $-$0.2$\pm$2.0 & 0.00$\pm$0.05 \\
0746 & 1525--1537 &   624&  6.1&  $+$1.0$\pm$1.0 & 0.16$\pm$0.16 &~& 1323 & 1495--1511 &   113&  8.6&  $-$0.4$\pm$1.0 & 0.04$\pm$0.11 \\
     &            &   606&  6.1&  $+$0.9$\pm$0.9 & 0.15$\pm$0.14 &~&      & 1518--1535 &   113& 10.7&  $-$0.7$\pm$1.3 & 0.07$\pm$0.12 \\
     &            &    18&  6.6&  $+$0.1$\pm$0.8 & 0.01$\pm$0.13 &~& 1332 & 1511--1530 &   503&  8.0&  $+$1.8$\pm$2.0 & 0.22$\pm$0.25 \\
     & 1539--1549 &   624&  8.3&  $+$0.4$\pm$0.9 & 0.05$\pm$0.11 &~& 1337 & 1528--1539 &   512&  5.1&  $-$0.1$\pm$0.6 & 0.03$\pm$0.11 \\
     &            &   606&  8.3&  $+$0.3$\pm$0.9 & 0.04$\pm$0.11 &~& 1359 & 1533--1549 &   933&  7.6&  $+$0.0$\pm$0.9 & 0.01$\pm$0.12 \\
     &            &    18&  8.5&  $+$0.1$\pm$0.9 & 0.01$\pm$0.11 &~& 1411 & 1534--1544 &   929&  3.1&  $-$1.8$\pm$0.6 & 0.57$\pm$0.20 \\
0824 & 1518--1539 &    87& 15.0&  $+$0.2$\pm$2.0 & 0.01$\pm$0.13 &~& 1413 & 1529--1549 &   465& 14.5&  $+$0.1$\pm$1.0 & 0.01$\pm$0.07 \\
0837 & 1502--1543 &   496& 33.2&  $-$0.3$\pm$0.9 & 0.01$\pm$0.03 &~& 1415 & 1476--1535 &   518& 32.6&  $-$4.0$\pm$0.8 & 0.12$\pm$0.03 \\
     &            &    21& 33.0&  $+$0.2$\pm$1.0 & 0.01$\pm$0.03 &~&      &            &   202& 30.7&  $-$7.9$\pm$1.0 & 0.26$\pm$0.03 \\
     &            &    12& 33.3&  $-$0.5$\pm$0.8 & 0.01$\pm$0.03 &~&      &            &   316& 28.7&  $+$3.9$\pm$0.8 & 0.14$\pm$0.03 \\
0839 & 1417--1461 &    17& 21.1&  $-$0.2$\pm$2.8 & 0.01$\pm$0.13 &~& 1441 & 1526--1542 &   792& 11.7&  $-$1.4$\pm$0.8 & 0.12$\pm$0.07 \\
     & 1497--1509 &    17&  7.4&  $-$0.2$\pm$1.4 & 0.02$\pm$0.19 &~& 1447 & 1447--1487 &   486&  7.6&  $+$1.2$\pm$0.8 & 0.15$\pm$0.10 \\
     & 1531--1554 &    17& 16.3&  $-$0.8$\pm$1.9 & 0.05$\pm$0.12 &~& 1516 & 1532--1552 &  1131& 11.7&  $-$2.0$\pm$0.8 & 0.17$\pm$0.07 \\
0844 & 1412--1529 &   420& 57.1&  $+$2.2$\pm$0.9 & 0.04$\pm$0.02 &~& 1542 & 1525--1554 &    17& 20.3&  $-$1.2$\pm$2.0 & 0.06$\pm$0.10 \\
0856 & 1508--1527 &   546& 11.2&  $-$0.7$\pm$0.6 & 0.06$\pm$0.05 &~& 1553 & 1476--1527 &   473& 21.0&  $-$3.4$\pm$1.8 & 0.16$\pm$0.09 \\
     &            &   520& 11.2&  $-$0.8$\pm$0.6 & 0.07$\pm$0.06 &~&      &            &   132& 21.0&  $-$3.2$\pm$2.0 & 0.15$\pm$0.10 \\
     &            &    26& 10.9&  $+$0.0$\pm$0.6 & 0.00$\pm$0.05 &~&      &            &   341& 19.3&  $-$0.2$\pm$1.5 & 0.01$\pm$0.08 \\
0903 & 1491--1515 &   319& 11.7&  $+$1.0$\pm$1.2 & 0.08$\pm$0.10 &~& 1609 & 1531--1547 &   588& 12.1&  $+$0.7$\pm$0.8 & 0.06$\pm$0.07 \\
0915 & 1535--1551 &   130&  4.8&  $-$0.4$\pm$1.0 & 0.08$\pm$0.20 &~& 1624 & 1403--1446 &   396& 14.8&  $-$1.1$\pm$0.7 & 0.07$\pm$0.05 \\
0934 & 1482--1499 &   300&  6.8&  $+$0.2$\pm$0.5 & 0.03$\pm$0.07 &~&      &            &   343& 14.8&  $-$1.1$\pm$0.7 & 0.08$\pm$0.05 \\
     & 1406--1440 &   300& 10.4&  $+$0.5$\pm$0.7 & 0.04$\pm$0.07 &~&      &            &    53& 14.2&  $+$0.1$\pm$0.6 & 0.00$\pm$0.05 \\
0945 & 1405--1500 &   672& 42.8&  $-$1.4$\pm$3.2 & 0.03$\pm$0.08 &~&      & 1528--1543 &   396&  5.8&  $+$0.0$\pm$0.6 & 0.01$\pm$0.11 \\
     &            &   664& 41.8&  $-$3.3$\pm$3.0 & 0.08$\pm$0.07 &~&      &            &   343&  5.5&  $-$0.7$\pm$0.6 & 0.12$\pm$0.11 \\
     &            &     8& 41.1&  $+$1.9$\pm$3.4 & 0.05$\pm$0.08 &~&      &            &    53&  5.5&  $+$0.7$\pm$0.6 & 0.12$\pm$0.10 \\
1007 & 1491--1532 &  6163&  4.7&  $-$2.0$\pm$0.7 & 0.43$\pm$0.16 &~& 2257 & 1537--1554 &  1318& 14.3&  $-$0.6$\pm$1.5 & 0.04$\pm$0.10 \\
     &            &  1095&  9.9&  $+$8.4$\pm$1.0 & 0.85$\pm$0.12 &~&      &            &   129& 14.4&  $-$0.4$\pm$1.7 & 0.03$\pm$0.12 \\
     &            &  5068&  8.9& $-$10.4$\pm$0.9 & 1.17$\pm$0.12 &~&      &            &  1189& 14.1&  $-$0.2$\pm$1.4 & 0.01$\pm$0.10 \\
     &            &      &     &                 &               &~& 2357 & 1525--1541 &   103& 12.8&  $-$0.0$\pm$0.9 & 0.00$\pm$0.07 \\
\hline                                              
\end{tabular}                                       
\end{minipage}                                      
\end{table*}                                        
\clearpage

\begin{table*}
\centering
\scriptsize
\begin{minipage}{17.5cm}
\caption{BAL variability in RQQs}
\begin{tabular}{p{60pt}rrrrrrp{60pt}rrrrrr}
\hline
Coordinates & $z$ & ${\Delta}\tau$\footnote{These and following
columns defined as in Table~1.} & ${\langle}EW{\rangle}$ &
${\Delta}EW$ & $\frac{|{\Delta}EW|}{{\langle}EW{\rangle}}$ & Ref & 
Coordinates & $z$ & ${\Delta}\tau$ & ${\langle}EW{\rangle}$ &
${\Delta}EW$ & $\frac{|{\Delta}EW|}{{\langle}EW{\rangle}}$ & Ref \\
\hline
  001130.5$+$005550 & 2.31  &    219 &  12.1 &  $-$0.7$\pm$1.4 & 0.06 & G09    & 123355.6$+$130409 & 2.38 &   2411 &  30.5 &  $+$6.2$\pm$1.7 & 0.20 &  G10 \\
  001219.6$+$023636 & 2.64  &   2118 &  21.3 &  $+$17.9$\pm$2.2 & 0.84 & G10   & 123458.1$+$130855 & 2.36 &   2038 &  60.6 &  $-$1.3$\pm$0.6 & 0.02 &  C11 \\
  001306.1$+$000431 & 2.16  &   1498 &   5.6 &  $-$3.5$\pm$1.5 & 0.63 & G10    & 123724.5$+$010615 & 2.02 &   1680 &   6.1 &  $-$0.7$\pm$1.0 & 0.12 &  G10 \\
  001927.8$+$003359 & 1.62  &    276 &  15.6 &  $-$6.3$\pm$1.4 & 0.41 & G09    & 123736.4$+$143640 & 2.70 &   1790 &  37.7 &  $+$4.2$\pm$3.2 & 0.11 &  G10 \\
  002127.8$+$010420 & 1.82  &   1571 &  12.4 &  $+$2.3$\pm$1.3 & 0.19 & G10    & 123754.8$+$084106 & 2.90 &   2082 &  25.0 &  $-$3.0$\pm$1.7 & 0.12 &  G10 \\
  002227.5$+$012413 & 2.13  &   2447 &  27.8 &  $+$11.8$\pm$1.5 & 0.43 & G10   & 124303.6$+$155047 & 2.36 &   2192 &  31.3 &  $+$6.3$\pm$1.7 & 0.20 &  G10 \\
  002410.8$-$015647 & 2.35  &   2118 &  45.3 &  $+$4.9$\pm$1.7 & 0.11 & G10    & 124551.4$+$010505 & 2.81 &   2009 &  39.3 &  $-$4.9$\pm$2.6 & 0.13 &  G10 \\
  002435.3$+$020648 & 2.83  &   2045 &  13.2 &  $-$5.6$\pm$1.9 & 0.43 & G10    & 124913.8$-$055919 & 2.25 &   1823 &  30.9 &  $-$0.4$\pm$0.5 & 0.01 &  C11 \\
  002733.8$-$013452 & 2.08  &   2484 &  21.5 &  $-$11.0$\pm$1.7 & 0.51 & G10   & 130058.1$+$010551 & 1.91 &    105 &  43.0 &  $+$2.6$\pm$1.3 & 0.06 &  L07 \\
  003135.5$+$003421 & 2.25  &   2338 &  35.0 &  $+$6.0$\pm$1.8 & 0.17 & G10    & 130554.7$+$303252 & 1.77 &   2104 &  39.9 &  $+$35.6$\pm$0.9 & 0.89 &  C11 \\
  003218.4$+$073832 & 3.25  &     89 &   7.2 &  $-$0.3$\pm$0.6 & 0.04 & B93    & 131136.5$-$055239 & 2.16 &   1684 &  58.7 &  $-$3.3$\pm$0.4 & 0.06 &  C11 \\
  004118.5$+$001742 & 1.77  &    170 &  11.2 &  $+$0.4$\pm$1.4 & 0.04 & G09    & 131305.7$+$015926 & 2.02 &     88 &   9.5 &  $+$1.6$\pm$0.4 & 0.17 &  L07 \\
  004527.6$+$143816 & 1.99  &     19 &  35.3 &  $-$0.4$\pm$0.3 & 0.01 & L07    & 131714.2$+$010013 & 2.70 &   2082 &  32.1 &  $+$20.0$\pm$2.5 & 0.62 &  G10 \\
  005355.1$-$000309 & 1.72  &   1461 &  34.8 &  $-$11.5$\pm$1.4 & 0.33 & G10   & 131853.4$+$002211 & 2.07 &    104 &  16.4 &  $+$1.3$\pm$0.6 & 0.08 &  L07 \\
  005824.7$+$004113 & 1.92  &   1498 &   5.2 &  $-$2.3$\pm$1.1 & 0.45 & G10    & 132742.9$+$003532 & 1.88 &    103 &   3.8 &  $+$1.5$\pm$0.6 & 0.39 &  L07 \\
  011227.6$-$011221 & 1.76  &   1461 &  21.0 &  $+$16.9$\pm$1.2 & 0.80 & G10   & 133901.8$+$132018 & 2.45 &   2016 &  42.2 &  $-$2.0$\pm$0.6 & 0.05 &  C11 \\
  011237.3$+$001929 & 2.66  &    523 &   4.0 &  $+$1.4$\pm$1.4 & 0.35 & G09    & 134544.5$+$002810 & 2.45 &     80 &  18.3 &  $+$2.8$\pm$0.5 & 0.15 &  L07 \\
  011913.2$+$005115 & 1.67  &    157 &  24.6 &  $+$4.4$\pm$1.4 & 0.18 & G09    & 142333.5$+$573909 & 1.87 &    547 &  18.8 &  $-$6.0$\pm$1.4 & 0.32 &  G09 \\
  012209.9$+$032544 & 2.09  &   2034 &  32.3 &  $-$3.3$\pm$1.6 & 0.10 & C11    & 142500.2$+$494729 & 2.25 &   1801 &  20.6 &  $-$3.8$\pm$1.2 & 0.18 &  C11 \\
  013625.6$-$103346 & 2.01  &    487 &  26.5 &  $-$2.1$\pm$1.4 & 0.08 & G09    & 143130.0$+$570138 & 1.80 &     31 &  16.8 &  $-$2.1$\pm$0.6 & 0.13 &  L07 \\
  014055.5$+$003908 & 1.49  &    167 &  10.0 &  $+$0.0$\pm$1.4 & 0.00 & G09    & 143641.2$+$001558 & 1.87 &     18 &  23.1 &  $+$0.6$\pm$0.7 & 0.03 &  L07 \\
  014817.5$+$043119 & 2.03  &    281 &  21.6 &  $+$3.2$\pm$2.0 & 0.15 & B93    & 143647.5$+$495256 & 1.59 &   2283 &  63.1 &  $-$6.9$\pm$2.0 & 0.11 &  C11 \\
  014918.7$+$015723 & 2.91  &   1695 &  42.7 &  $-$9.8$\pm$0.5 & 0.23 & C11    & 143907.5$-$010616 & 1.82 &    265 &   7.2 &  $+$0.7$\pm$1.4 & 0.10 &  G09 \\
  020006.3$-$003709 & 2.14  &    575 &  52.3 &  $+$2.5$\pm$1.4 & 0.05 & G09    & 144244.2$-$010943 & 1.93 &    256 &   9.1 &  $+$0.1$\pm$1.4 & 0.01 &  G09 \\
  022839.2$-$101111 & 2.26  &   1702 &  48.8 &  $+$5.5$\pm$0.3 & 0.11 & C11    & 144403.9$+$565751 & 1.86 &     31 &   1.6 &  $+$0.5$\pm$0.3 & 0.32 &  L07 \\
  024413.7$-$000447 & 2.80  &    210 &  18.9 &  $-$5.9$\pm$1.4 & 0.31 & G09    & 144514.8$-$002358 & 2.24 &   2301 &  38.5 &  $+$2.5$\pm$1.5 & 0.07 &  G10 \\
  024701.1$+$000330 & 2.15  &    350 &   9.1 &  $-$0.2$\pm$1.4 & 0.02 & G09    & 144545.3$+$012912 & 2.45 &   2155 &  44.4 &  $+$0.9$\pm$1.7 & 0.02 &  G10 \\
  025042.4$+$003536 & 2.39  &    342 &  47.3 &  $+$0.2$\pm$1.4 & 0.00 & G09    & 144959.9$+$003225 & 1.72 &    121 &  10.5 &  $-$0.2$\pm$0.6 & 0.02 &  L07 \\
  025751.5$+$002045 & 1.51  &    155 &  11.1 &  $+$1.2$\pm$1.4 & 0.11 & G09    & 145428.5$+$571441 & 3.27 &     21 &   3.9 &  $+$0.9$\pm$0.5 & 0.22 &  L07 \\
  030504.9$+$171653 & 2.89  &   1614 &  12.9 &  $+$0.1$\pm$0.3 & 0.00 & C11    & 150109.1$-$011502 & 2.13 &    120 &  23.2 &  $-$7.8$\pm$1.5 & 0.34 &  L07 \\
  031828.9$-$001523 & 1.98  &     36 &   3.1 &  $+$1.5$\pm$0.5 & 0.48 & L07    & 151927.4$-$010729 & 1.72 &    265 &  11.5 &  $+$0.4$\pm$1.4 & 0.03 &  G09 \\
  075010.1$+$304032 & 1.89  &    109 &   3.2 &  $-$4.7$\pm$0.3 & 1.45 & L07    & 152553.8$+$513649 & 2.88 &   1578 &  16.7 &  $-$2.6$\pm$0.2 & 0.16 &  C11 \\
  081416.7$+$435405 & 1.35  &     92 &  16.7 &  $+$2.4$\pm$1.4 & 0.14 & L07    & 160649.2$+$451051 & 2.83 &     23 &   4.6 &  $+$1.1$\pm$0.7 & 0.23 &  L07 \\
  081657.5$+$060441 & 2.01  &     75 &  11.5 &  $-$5.1$\pm$1.7 & 0.44 & L07    & 170114.3$+$222448 & 1.90 &    199 &   3.7 &  $+$1.2$\pm$1.4 & 0.32 &  G09 \\
  081822.6$+$434633 & 2.04  &     81 &   1.5 &  $+$2.7$\pm$0.3 & 1.82 & L07    & 170633.0$+$615715 & 2.01 &     61 &   4.0 &  $+$0.8$\pm$0.3 & 0.20 &  L07 \\
  084538.6$+$342043 & 2.15  &   2023 &  24.5 &  $-$0.4$\pm$0.7 & 0.02 & C11    & 172001.3$+$621245 & 1.76 &     34 &  23.0 &  $-$0.1$\pm$1.4 & 0.01 &  L07 \\
  084908.1$+$152931 & 2.93  &   1381 &   6.3 &  $+$1.6$\pm$0.3 & 0.26 & C11    & 222830.4$-$051855 & 1.98 &    375 &  45.1 &  $+$4.8$\pm$1.9 & 0.11 &  B93 \\
  090638.1$+$172223 & 2.69  &   1841 &  66.0 &  $-$17.4$\pm$0.5 & 0.26 & C11   & 224019.0$+$144435 & 2.24 &     79 &   1.9 &  $-$0.1$\pm$0.6 & 0.08 &  L07 \\
  093552.9$+$495314 & 1.93  &   2301 &  49.4 &  $+$1.7$\pm$0.5 & 0.04 & C11    & 224320.4$-$004918 & 1.78 &    160 &  17.5 &  $+$9.7$\pm$1.4 & 0.55 &  G09 \\
  093620.5$+$004649 & 1.72  &    106 &   3.4 &  $+$5.2$\pm$0.7 & 1.50 & L07    & 224324.5$-$005330 & 1.90 &    154 &  22.2 &  $-$5.6$\pm$1.4 & 0.25 &  G09 \\
  094425.4$+$610934 & 2.27  &    569 &  36.8 &  $-$2.8$\pm$1.4 & 0.08 & G09    & 224733.4$-$002009 & 2.51 &    309 &  23.9 &  $+$1.8$\pm$1.4 & 0.08 &  G09 \\
  095933.6$-$054952 & 1.81  &   2082 &  18.8 &  $+$0.1$\pm$0.7 & 0.01 & C11    & 225746.9$+$000121 & 1.73 &    298 &   5.8 &  $+$1.5$\pm$1.4 & 0.26 &  G09 \\
  101341.8$+$085126 & 2.27  &   1804 &  48.4 &  $-$0.2$\pm$1.1 & 0.01 & C11    & 234315.8$+$004659 & 2.78 &    392 &  10.1 &  $+$0.7$\pm$1.4 & 0.07 &  G09 \\
  102250.1$+$483631 & 2.07  &    107 &  27.7 &  $-$2.7$\pm$0.6 & 0.10 & L07    & 234506.3$+$010135 & 1.80 &    231 &  32.2 &  $+$2.9$\pm$1.4 & 0.09 &  G09 \\
  112820.8$-$011441 & 1.80  &    272 &  13.8 &  $+$1.3$\pm$1.4 & 0.09 & G09    & 235224.1$-$000951 & 2.74 &    197 &  20.7 &  $+$3.4$\pm$1.4 & 0.16 &  G09 \\
  115031.0$-$004403 & 2.39  &     83 &  11.6 &  $+$0.3$\pm$1.6 & 0.03 & L07    & 235253.5$-$002850 & 1.62 &    281 &  30.1 &  $+$3.4$\pm$1.4 & 0.11 &  G09 \\
  121125.4$+$151851 & 1.96  &   2228 &  28.7 &  $+$6.4$\pm$1.6 & 0.22 & G10    & 235408.6$-$001615 & 1.77 &    265 &  41.8 &  $-$12.1$\pm$1.4 & 0.29 &  G09 \\
\hline                                              
\end{tabular}                                       
\end{minipage}                                      
\end{table*}                                        
\clearpage
																				      
\begin{table*}
\centering
\scriptsize
\begin{minipage}{17.0cm}
\caption{BAL variability in RLQs and RQQs}
\begin{tabular}{p{60pt}rrrrrrrrrrrrrrrr}
\hline
& & \multicolumn{3}{c}{${\Delta}\tau$ (d)} & & \multicolumn{3}{c}{${\langle}EW{\rangle}$ (\AA)} & & \multicolumn{3}{c}{$|{\Delta}EW|$ (\AA)} & 
& \multicolumn{3}{c}{$|{\Delta}EW|/{\langle}EW{\rangle}$} \\
Sample & $n$ & Med & Mean & KS $p$ & & Med & Mean & KS $p$ & & Med & Mean & KS $p$ & & Med & Mean & KS $p$ \\
\hline
\multicolumn{17}{c}{Groupings of RLQs} \\
\hline
All values                  &   78  &   473 &    665$\pm$108 &   \nodata  &&   14.5  &   20.7$\pm$1.6  &  \nodata  &&   1.0  &    1.9$\pm$0.3  &  \nodata  && 0.06  &   0.12$\pm$0.02  &  \nodata  \\ 
Longest sep                 &   46  &   503 &    724$\pm$143 &   \nodata  &&   12.8  &   18.6$\pm$2.2  &  \nodata  &&   1.0  &    1.7$\pm$0.4  &  \nodata  && 0.06  &   0.12$\pm$0.02  &  \nodata  \\ [+4pt]
~~~Core dom                 &   39  &   465 &    579$\pm$ 87 &   \nodata  &&   14.5  &   19.9$\pm$2.4  &  \nodata  &&   0.7  &    1.7$\pm$0.5  &  \nodata  && 0.04  &   0.09$\pm$0.02  &  \nodata  \\ 
~~~Lobe dom                 &    7  &   770 &   1532$\pm$778 &    0.04    &&    7.8  &   11.4$\pm$5.0  &   0.01    &&   1.8  &    1.8$\pm$0.3  &   0.05    && 0.22  &   0.24$\pm$0.07  &   0.04    \\ [+4pt]
~~~${\ell}_{\rm r}<33$      &   24  &   319 &    726$\pm$262 &   \nodata  &&   15.4  &   21.1$\pm$3.2  &  \nodata  &&   1.1  &    1.6$\pm$0.4  &  \nodata  && 0.06  &   0.09$\pm$0.02  &  \nodata  \\ 
~~~${\ell}_{\rm r}\ge33$    &   22  &   624 &    723$\pm$ 95 &    0.01    &&   11.6  &   15.8$\pm$3.0  &   0.21    &&   1.0  &    1.8$\pm$0.7  &   1.00    && 0.05  &   0.15$\pm$0.04  &   0.96    \\ [+4pt]
~~~$R^{*}<2$                &   30  &   419 &    519$\pm$106 &   \nodata  &&   15.0  &   20.8$\pm$3.1  &  \nodata  &&   1.0  &    1.8$\pm$0.6  &  \nodata  && 0.04  &   0.10$\pm$0.03  &  \nodata  \\ 
~~~$R^{*}\ge2$              &   16  &   745 &   1110$\pm$346 &    0.00    &&   11.6  &   14.4$\pm$2.8  &   0.28    &&   1.4  &    1.4$\pm$0.3  &   0.54    && 0.06  &   0.15$\pm$0.04  &   0.52    \\ 
\hline
\multicolumn{17}{c}{Groupings of RQQs} \\
\hline
All values                  &  115  &   281 &    808$\pm$ 80 &   \nodata  &&   21.6  &   24.7$\pm$1.5  &  \nodata  &&   2.1  &    3.6$\pm$0.5  &  \nodata  && 0.11  &   0.21$\pm$0.03  &  \nodata  \\ 
Longest sep                 &   94  &   350 &    921$\pm$ 92 &   \nodata  &&   21.0  &   23.2$\pm$1.7  &  \nodata  &&   2.5  &    4.1$\pm$0.5  &  \nodata  && 0.13  &   0.24$\pm$0.03  &  \nodata  \\ [+4pt]
~~~Lundgren+07              &   24  &    81 &     72$\pm$  7 &    0.00    &&   11.5  &   13.2$\pm$2.3  &   0.01    &&   1.5  &    2.0$\pm$0.4  &   0.20    && 0.17  &   0.35$\pm$0.10  &   0.69    \\ 
~~~Gibson+09                &   28  &   265 &    296$\pm$ 25 &    0.00    &&   17.5  &   19.6$\pm$2.5  &   0.38    &&   1.8  &    2.8$\pm$0.6  &   1.00    && 0.10  &   0.16$\pm$0.03  &   0.76    \\ 
~~~Capellupo+11             &   18  &  1841 &   1878$\pm$ 59 &    0.00    &&   42.2  &   37.9$\pm$4.3  &   0.04    &&   2.6  &    5.3$\pm$2.1  &   0.98    && 0.10  &   0.14$\pm$0.05  &   0.53    \\ 
~~~Gibson+10                &   21  &  2082 &   1999$\pm$ 75 &    0.00    &&   28.7  &   26.5$\pm$2.7  &   0.26    &&   5.6  &    7.1$\pm$1.2  &   0.00    && 0.20  &   0.32$\pm$0.05  &   0.02    \\ 
\hline
\multicolumn{17}{c}{Comparison of RLQs to RQQs (for $EW>3.5$~\AA, longest separation)}\\
\hline
{\bf RLQs                        }&{\bf   42  }&{\bf   503 }&{\bf    740$\pm$155 }&   \nodata  &&{\bf   14.5  }&{\bf   20.1$\pm$2.3  }&  \nodata  &&{\bf   1.0  }&{\bf    1.8$\pm$0.5  }&  \nodata  &&{\bf 0.05  }&{\bf   0.10$\pm$0.02  }&  \nodata  \\ 
RQQs                        &   88  &   487 &    978$\pm$ 96 &    0.00    &&   21.6  &   24.6$\pm$1.7  &   0.02    &&   2.5  &    4.2$\pm$0.6  &   0.00    && 0.12  &   0.19$\pm$0.02  &   0.00    \\ 
{\bf RQQs matched                }&{\bf   42  }&{\bf   375 }&{\bf    738$\pm$106 }&{\bf    0.26    }&&{\bf   18.8  }&{\bf   22.1$\pm$2.3  }&{\bf   0.56    }&&{\bf   2.3  }&{\bf    3.3$\pm$0.6  }&{\bf   0.01    }&&{\bf 0.11  }&{\bf   0.17$\pm$0.03  }&{\bf   0.01  }  \\ [+4pt]
				    														       
RLQs $EW<20$~\AA            &   28  &   512 &    736$\pm$214 &   \nodata  &&   11.7  &   11.2$\pm$0.7  &  \nodata  &&   0.7  &    1.1$\pm$0.2  &  \nodata  && 0.06  &   0.10$\pm$0.02  &  \nodata  \\ 
RQQs $EW<20$~\AA            &   39  &   219 &    529$\pm$104 &    0.01    &&   11.1  &   10.8$\pm$0.8  &   0.33    &&   1.3  &    2.0$\pm$0.3  &   0.17    && 0.15  &   0.19$\pm$0.03  &   0.07    \\ 
RLQs $EW\ge20$~\AA          &   14  &   496 &    749$\pm$194 &   \nodata  &&   39.9  &   37.9$\pm$3.4  &  \nodata  &&   1.4  &    3.2$\pm$1.2  &  \nodata  && 0.04  &   0.09$\pm$0.04  &  \nodata  \\ 
RQQs $EW\ge20$~\AA          &   49  &  1790 &   1336$\pm$130 &    0.05    &&   32.3  &   35.6$\pm$1.7  &   0.58    &&   3.4  &    5.9$\pm$0.9  &   0.00    && 0.11  &   0.19$\pm$0.03  &   0.01    \\ [+4pt]
RLQs ${\Delta}\tau<500$~d   &   21  &   130 &    233$\pm$ 39 &   \nodata  &&   16.2  &   21.2$\pm$3.3  &  \nodata  &&   0.8  &    1.0$\pm$0.2  &  \nodata  && 0.04  &   0.06$\pm$0.02  &  \nodata  \\ 
RQQs ${\Delta}\tau<500$~d   &   45  &   160 &    180$\pm$ 17 &    0.10    &&   16.4  &   18.1$\pm$1.7  &   0.67    &&   1.5  &    2.4$\pm$0.4  &   0.03    && 0.11  &   0.15$\pm$0.02  &   0.01    \\ 
RLQs ${\Delta}\tau\ge500$~d &   21  &   854 &   1247$\pm$268 &   \nodata  &&   11.7  &   19.0$\pm$3.3  &  \nodata  &&   1.4  &    2.5$\pm$0.9  &  \nodata  && 0.06  &   0.13$\pm$0.03  &  \nodata  \\ 
RQQs ${\Delta}\tau\ge500$~d &   43  &  2009 &   1814$\pm$ 77 &    0.00    &&   31.3  &   31.4$\pm$2.5  &   0.00    &&   3.5  &    6.0$\pm$1.1  &   0.00    && 0.16  &   0.23$\pm$0.04  &   0.10    \\ 
\hline								  									    
\end{tabular}   
~\\
{\it Note\/}:~~The median and mean values are given for each grouping; the error on
the mean is $\sigma/\sqrt{n}$. The KS-test probability is versus the
nearest preceding grouping without entered values, and $p<0.005$
($p\ge0.995$) are represented as 0.00 (1.00).
\end{minipage}                                      
\end{table*}                                        
\clearpage

\begin{table*}
\centering
\scriptsize
\begin{minipage}{14.0cm}
\caption{Correlation tests}
\begin{tabular}{p{100pt}rrrrrrrrrrr}
\hline
 & & & \multicolumn{4}{c}{$|{\Delta}EW|$} & & \multicolumn{4}{c}{$|{\Delta}EW|/{\langle}EW{\rangle}$} \\
Sample & $n$ & & $p_{\tau}$ & $p_{\rho}$ & $\rho$ & $\beta$ & & $p_{\tau}$ & $p_{\rho}$ & $\rho$ & $\beta$ \\
\hline
\multicolumn{12}{c}{Versus ${\Delta}\tau$} \\
\hline
RLQs, all values            &  78  & &   0.99  &  0.99 &     0.30 &     0.47$\pm$0.23  & &   0.99  &  0.99 &     0.31 &     0.36$\pm$0.14 \\ 
RLQs, longest sep           &  46  & &   0.97  &  0.97 &     0.32 &     0.50$\pm$0.28  & &   0.96  &  0.95 &     0.29 &     0.72$\pm$0.28 \\ 
RLQs, $EW>3.5$\AA           &  42  & &   0.97  &  0.97 &     0.34 &     0.53$\pm$0.29  & &   0.97  &  0.96 &     0.32 &     0.68$\pm$0.26 \\ 
RLQs, core dom              &  39  & &   0.81  &  0.81 &     0.21 &     0.24$\pm$0.29  & &   0.81  &  0.79 &     0.20 &     0.14$\pm$0.16 \\ 
RLQs, lobe dom              &   7  & &   0.64  &  0.60 &     0.38 &     1.46$\pm$0.71  & &   0.35  &  0.30 &     0.18 &     1.77$\pm$2.03 \\ 
RQQs, all values            & 115  & &   1.00  &  1.00 &     0.36 &     1.55$\pm$0.40  & &   0.44  &  0.42 &     0.05 &     0.39$\pm$0.24 \\ 
RQQs, longest sep           &  94  & &   1.00  &  1.00 &     0.40 &     1.81$\pm$0.48  & &   0.03  &  0.03 &     0.00 &     0.31$\pm$0.28 \\ 
RQQs, $EW>3.5$\AA           &  88  & &   1.00  &  1.00 &     0.39 &     1.93$\pm$0.53  & &   0.67  &  0.67 &     0.10 &     0.45$\pm$0.28 \\ 
RQQs, matched               &  42  & &   0.99  &  0.99 &     0.40 &     1.47$\pm$0.63  & &   0.96  &  0.96 &     0.32 &     1.04$\pm$0.40 \\ 
RQQs+RLQs $EW>3.5$\AA       & 130  & &   1.00  &  1.00 &     0.39 &     1.29$\pm$0.34  & &   0.97  &  0.97 &     0.19 &     0.52$\pm$0.21 \\ 
\hline
\multicolumn{12}{c}{Versus ${\langle}EW{\rangle}$} \\                                         
\hline
RLQs, all values            &  78  & &   1.00  &  1.00 &     0.34 &     0.29$\pm$0.10  & &   0.87  &  0.90 &  $-$0.19 &  $-$0.01$\pm$0.01 \\ 
RLQs, longest sep           &  46  & &   0.98  &  0.98 &     0.34 &     0.15$\pm$0.11  & &   0.93  &  0.92 &  $-$0.26 &  $-$0.01$\pm$0.01 \\ 
RLQs, $EW>3.5$\AA           &  42  & &   0.97  &  0.97 &     0.34 &     0.14$\pm$0.12  & &   0.75  &  0.71 &  $-$0.17 &  $-$0.01$\pm$0.01 \\ 
RLQs, core dom              &  39  & &   1.00  &  1.00 &     0.47 &     0.23$\pm$0.08  & &   0.51  &  0.51 &  $-$0.11 &     0.00$\pm$0.01 \\ 
RLQs, lobe dom              &   7  & &   0.64  &  0.64 &     0.41 &     0.06$\pm$0.28  & &   0.35  &  0.36 &  $-$0.21 &  $-$0.07$\pm$0.05 \\ 
RQQs, all values            & 115  & &   1.00  &  1.00 &     0.30 &     0.41$\pm$0.15  & &   1.00  &  1.00 &  $-$0.34 &  $-$0.03$\pm$0.01 \\ 
RQQs, longest sep           &  94  & &   1.00  &  1.00 &     0.40 &     0.71$\pm$0.19  & &   1.00  &  1.00 &  $-$0.29 &  $-$0.02$\pm$0.01 \\ 
RQQs, $EW>3.5$\AA           &  88  & &   1.00  &  1.00 &     0.42 &     0.76$\pm$0.21  & &   0.94  &  0.94 &  $-$0.20 &  $-$0.02$\pm$0.01 \\ 
RQQs, matched               &  42  & &   0.98  &  0.99 &     0.38 &     0.74$\pm$0.24  & &   0.83  &  0.81 &  $-$0.20 &  $-$0.02$\pm$0.02 \\ 
RQQs+RLQs $EW>3.5$\AA       & 130  & &   1.00  &  1.00 &     0.43 &     0.65$\pm$0.13  & &   0.83  &  0.82 &  $-$0.12 &  $-$0.02$\pm$0.01 \\ 
\hline
\multicolumn{12}{c}{Versus optical continuum variability} \\                                         
\hline
RLQs, longest sep           &  44  & &   0.25  &  0.14 &     0.03 &     0.11$\pm$0.21  & &   0.69  &  0.63 &     0.14 &     0.17$\pm$0.17 \\ 
RLQs, core dom              &  39  & &   0.17  &  0.08 &     0.15 &     0.15$\pm$0.26  & &   0.81  &  0.74 &     0.19 &     0.35$\pm$0.14 \\ 
RLQs, lobe dom              &   7  & &   0.14  &  0.10 &  $-$0.06 &  $-$0.03$\pm$0.35  & &   0.41  &  0.42 &  $-$0.26 &  $-$0.62$\pm$0.68 \\ 
RQQs, longest sep           &  90  & &   0.11  &  0.09 &     0.01 &     0.00$\pm$0.53  & &   0.71  &  0.71 &     0.11 &     0.10$\pm$0.29 \\ 
RQQs, matched               &  45  & &   0.53  &  0.50 &     0.10 &  $-$0.10$\pm$0.62  & &   0.48  &  0.49 &     0.10 &     0.26$\pm$0.41 \\ 
RQQs+RLQs                   & 134  & &   0.59  &  0.54 &     0.06 &  $-$0.01$\pm$0.32  & &   0.92  &  0.91 &     0.15 &     0.12$\pm$0.20 \\ 
\hline
\multicolumn{12}{c}{Versus ${\ell}_{\rm r}$ (RLQs only)} \\                                         
\hline
RLQs, longest sep           &  46  & &   0.21  &  0.21 &  $-$0.04 &     0.00$\pm$0.24  & &   0.14  &  0.14 &     0.03 &     0.14$\pm$0.24 \\ 
RLQs, $EW>3.5$\AA           &  42  & &   0.13  &  0.14 &  $-$0.03 &     0.04$\pm$0.25  & &   0.07  &  0.07 &  $-$0.01 &     0.06$\pm$0.21 \\ 
RLQs, $3.5<EW<20$\AA        &  28  & &   0.40  &  0.28 &     0.07 &     0.27$\pm$0.29  & &   0.46  &  0.43 &     0.11 &     0.27$\pm$0.29 \\ 
RLQs, $EW\ge20$\AA          &  14  & &   0.91  &  0.86 &  $-$0.42 &  $-$0.55$\pm$0.39  & &   0.68  &  0.64 &  $-$0.27 &  $-$0.32$\pm$0.20 \\ 
RLQs, ${\Delta}\tau<500$~d  &  21  & &   0.10  &  0.08 &  $-$0.02 &  $-$0.15$\pm$0.31  & &   0.67  &  0.66 &  $-$0.22 &  $-$0.22$\pm$0.20 \\ 
RLQs, ${\Delta}\tau\ge500$~d&  21  & &   0.36  &  0.35 &  $-$0.11 &     0.09$\pm$0.40  & &   0.02  &  0.02 &     0.01 &     0.09$\pm$0.45 \\ 	      	     
\hline
\multicolumn{12}{c}{Versus $R^{*}$ (RLQs only)} \\
\hline
RLQs, longest sep           &  46  & &   0.32  &  0.25 &  $-$0.05 &     0.17$\pm$0.27  & &   0.45  &  0.49 &     0.10 &     0.35$\pm$0.27 \\ 
RLQs, $EW>3.5$\AA           &  42  & &   0.33  &  0.29 &  $-$0.06 &     0.21$\pm$0.30  & &   0.03  &  0.09 &     0.02 &     0.22$\pm$0.25 \\ 
RLQs, $3.5<EW<20$\AA        &  28  & &   0.77  &  0.74 &     0.22 &     0.66$\pm$0.33  & &   0.80  &  0.81 &     0.25 &     0.68$\pm$0.35 \\ 
RLQs, $EW\ge20$\AA          &  14  & &   0.99  &  0.99 &  $-$0.70 &  $-$0.54$\pm$0.41  & &   0.92  &  0.92 &  $-$0.48 &  $-$0.29$\pm$0.21 \\ 
RLQs, ${\Delta}\tau<500$~d  &  21  & &   0.55  &  0.44 &  $-$0.13 &  $-$0.28$\pm$0.43  & &   0.59  &  0.52 &  $-$0.16 &  $-$0.21$\pm$0.26 \\ 
RLQs, ${\Delta}\tau\ge500$~d&  21  & &   0.46  &  0.43 &  $-$0.13 &     0.31$\pm$0.46  & &   0.10  &  0.10 &  $-$0.03 &  $-$0.00$\pm$0.47 \\ 		
\hline
\end{tabular}   
~\\
{\it Note\/}:~~The probabilities $p_{\tau}$ and $p_{\rho}$ are for the Kendall's and Spearman tests, 
and $\rho$ is the Spearman correlation coefficient. The slope $\beta$ is calculated using the {\tt IDL}
routine robust\_linefit and for the linear regression ${\Delta}\tau$ is expressed as a logarithm, 
$|{\Delta}EW|/{\langle}EW{\rangle}$ is multiplied by 10 for all fits, and $|{\Delta}EW|$ is multiplied 
by 10 only versus ${\langle}EW{\rangle}$.
\end{minipage}                                      
\end{table*}                                        
\clearpage

\appendix

\begin{figure*}
\includegraphics[scale=0.9]{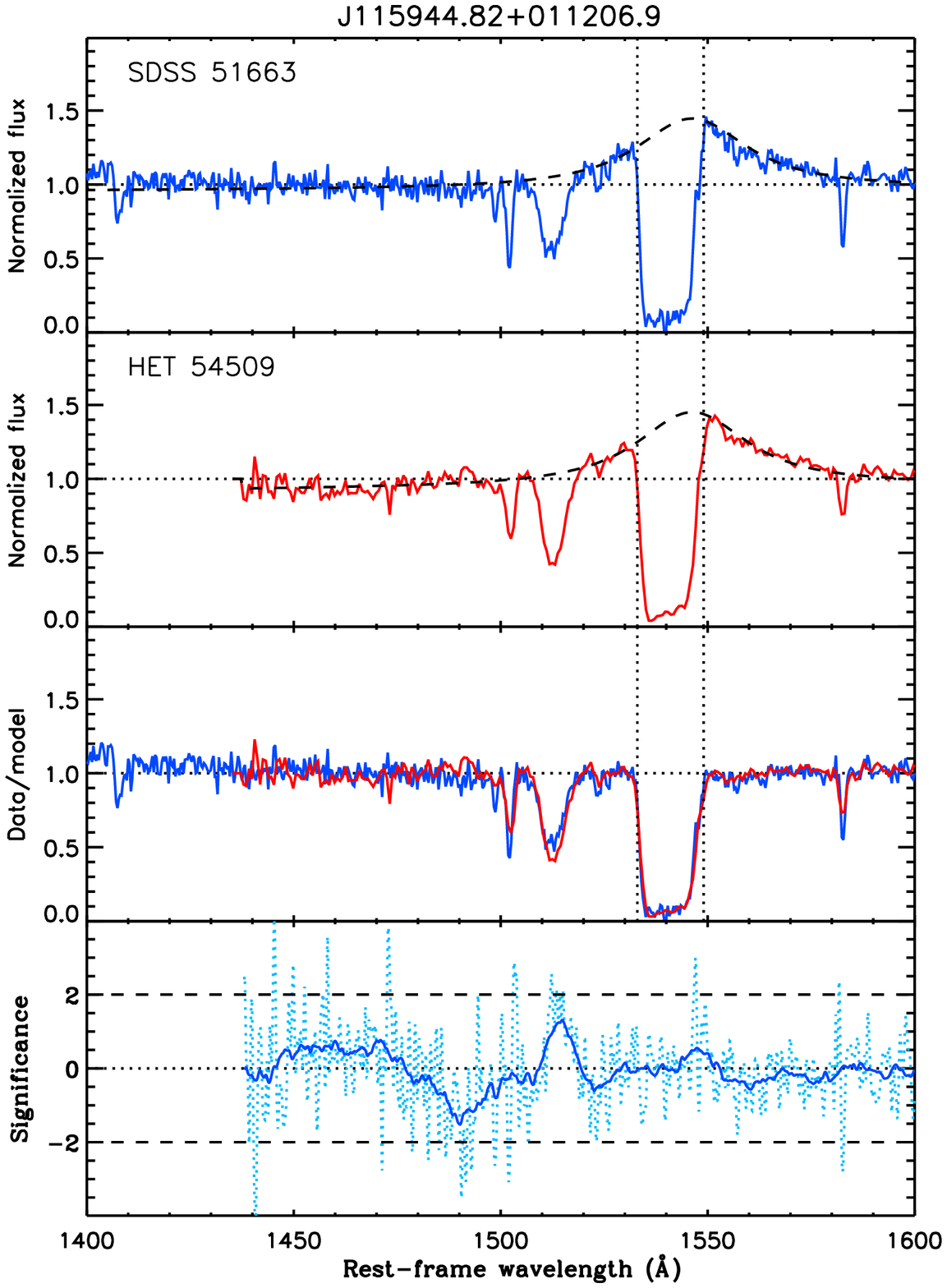}
\caption{\small Appendix A: Assessment of BAL variability for 115944.82$+$011206.9, as in Figure~4. (Selected example, full Appendix A has all BAL RLQs spectra used in this work.)}
\end{figure*}

\begin{figure*}
\includegraphics[scale=0.9]{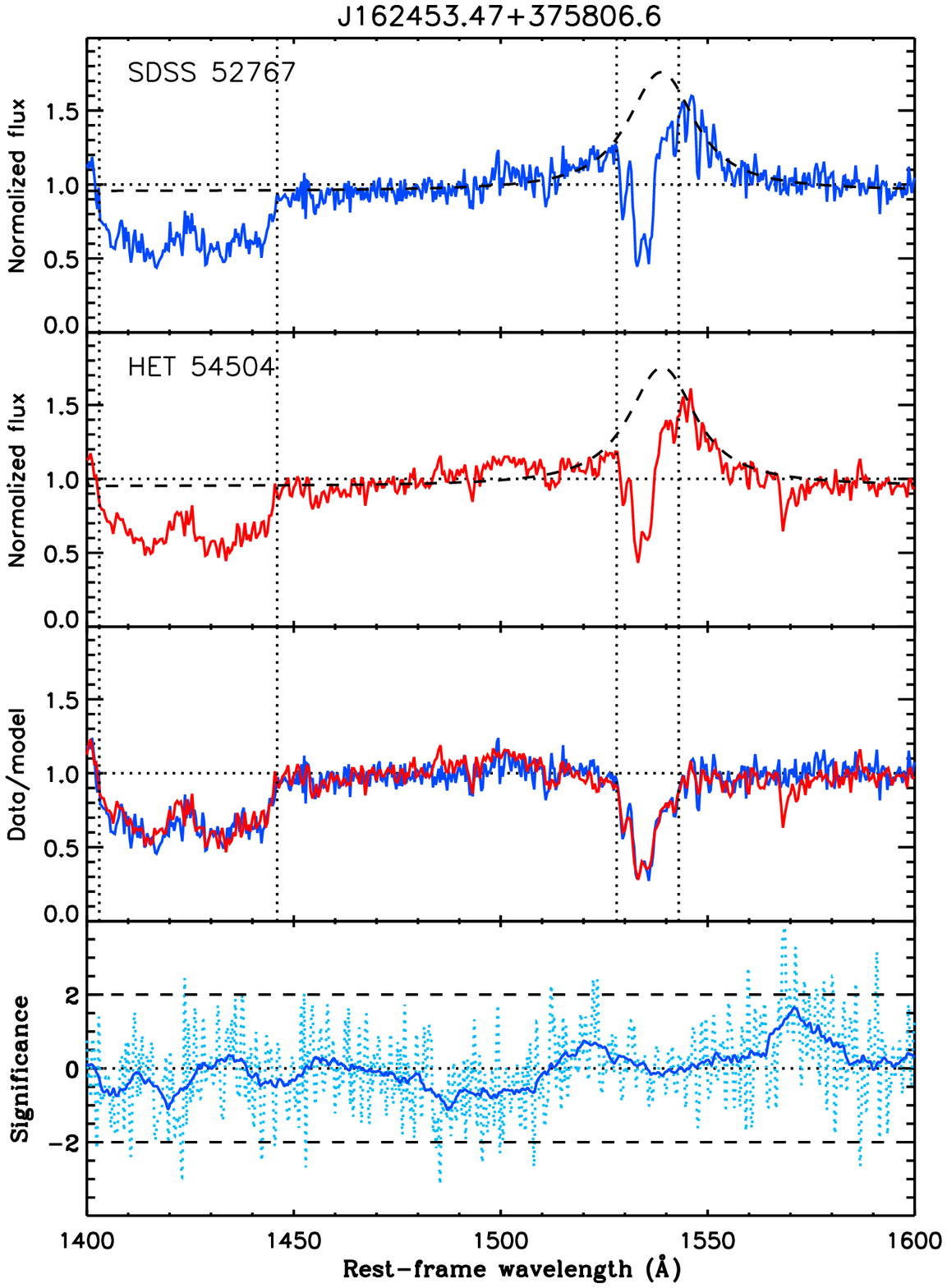}
\caption{\small Appendix A: Assessment of BAL variability for 162453.47$+$375806.6, as in Figure~4. (Selected example, full Appendix A has all BAL RLQs spectra used in this work.)}
\end{figure*}

\appendix

\begin{figure*}
\includegraphics[scale=0.75]{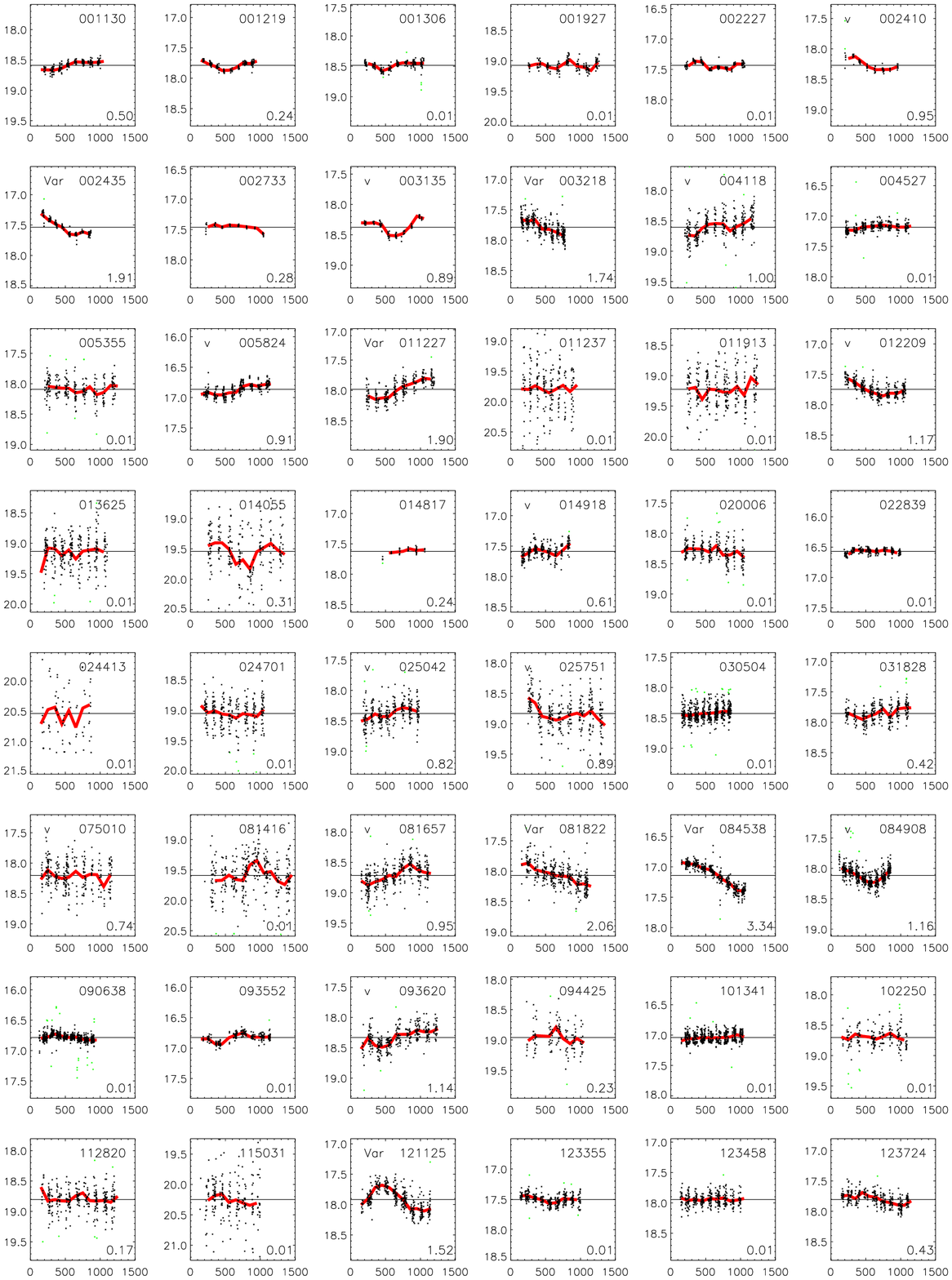}
\caption{\small Appendix B: Optical continuum variability for first set of BAL RQQs, as in Figure~13.}
\end{figure*}

\begin{figure*}
\includegraphics[scale=0.75]{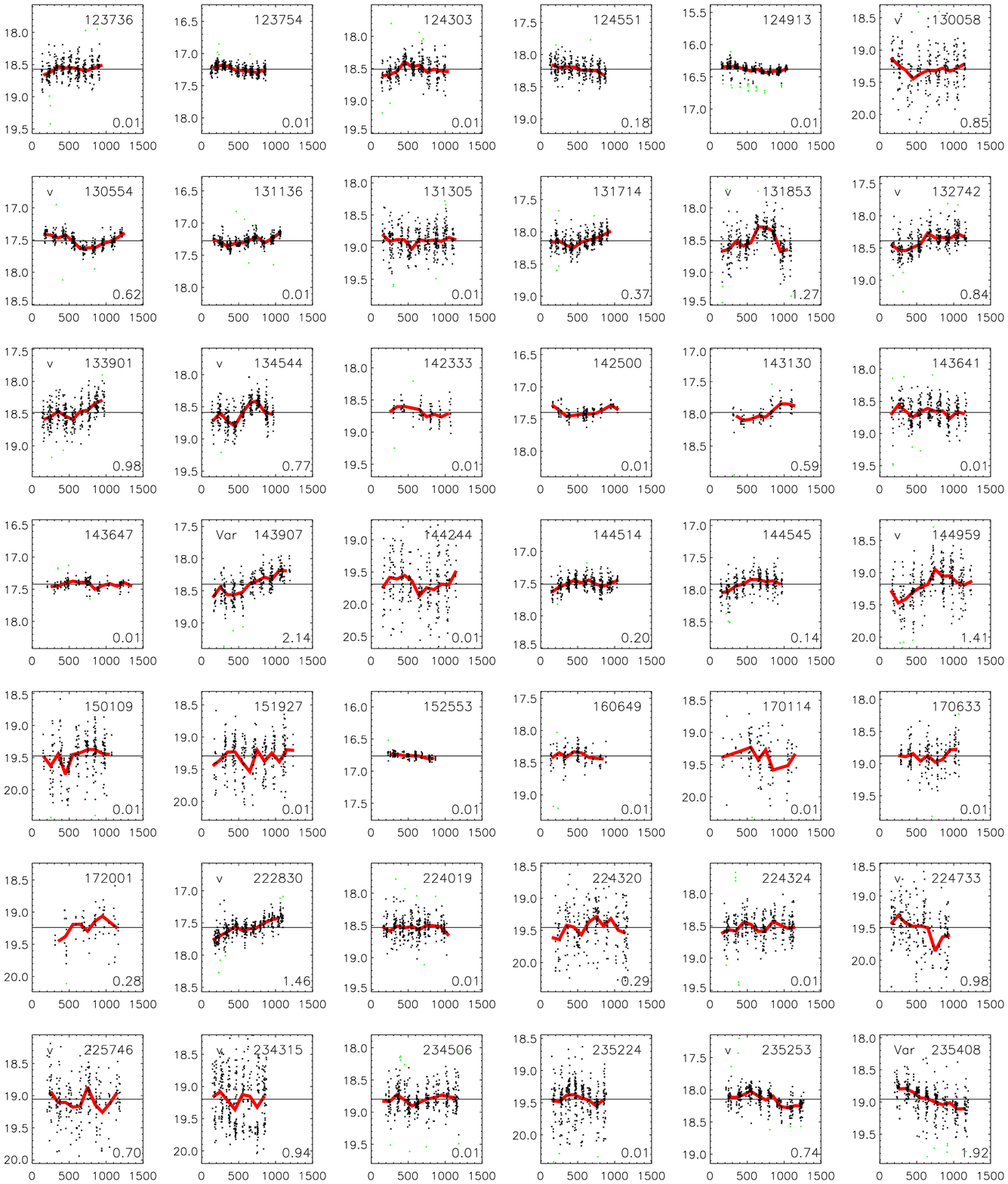} 
\caption{\small Appendix B: Optical continuum variability for second set of BAL RQQs, as in Figure~13.}
\end{figure*}

\end{document}